\documentclass[aps,prx,twocolumn,superscriptaddress,amsmath,amssymb,showpacs]{revtex4-2}

\usepackage{color}
\usepackage[dvipsnames]{xcolor}
\usepackage{graphicx}	
\usepackage{dcolumn}	
\usepackage{bm}		
\usepackage{verbatim} 	
\usepackage{braket}
\usepackage{xr}
\usepackage{hyperref}
\usepackage{booktabs}
\usepackage{upgreek}
\usepackage{notes2bib}
\usepackage{pdfpages}
\usepackage{lipsum}  
\usepackage{moresize}
\usepackage{empheq}
\usepackage{cancel}
\usepackage{physics}

\makeatletter
\AtBeginDocument{\let\LS@rot\@undefined}
\makeatother

\renewcommand{\d}{{\rm d}}

\newcolumntype{L}[1]{>{\raggedright\let\newline\\\arraybackslash\hspace{0pt}}m{#1}}
\newcolumntype{C}[1]{>{\centering\let\newline\\\arraybackslash\hspace{0pt}}m{#1}}
\newcolumntype{R}[1]{>{\raggedleft\let\newline\\\arraybackslash\hspace{0pt}}m{#1}}

\newcommand\dV{{\rm d}^3\vec{r}}

\renewcommand{\d}{{\rm d}}
\newcommand{\appropto}{\mathrel{\vcenter{
  \offinterlineskip\halign{\hfil$##$\cr
    \propto\cr\noalign{\kern2pt}\sim\cr\noalign{\kern-2pt}}}}}
\newcommand{\resim}{\mathord{\sim}}

\begin{document}

\title{A full degree-of-freedom photonic crystal spatial light modulator}

\author{Christopher L. Panuski}
\email{cpanuski@mit.edu}
\affiliation{Research Laboratory of Electronics, Massachusetts Institute of Technology, Cambridge, MA 02139, USA}

\author{Ian R. Christen}
\affiliation{Research Laboratory of Electronics, Massachusetts Institute of Technology, Cambridge, MA 02139, USA}

\author{Momchil Minkov}
\affiliation{Flexcompute, Inc., 130 Trapelo Rd., Belmont, MA 02478, USA}

\author{Cole J. Brabec}
\affiliation{Research Laboratory of Electronics, Massachusetts Institute of Technology, Cambridge, MA 02139, USA}

\author{Sivan Trajtenberg-Mills}
\affiliation{Research Laboratory of Electronics, Massachusetts Institute of Technology, Cambridge, MA 02139, USA}

\author{Alexander D. Griffiths}
\affiliation{Institute of Photonics, Dept. of Physics, University of Strathclyde, Technology and Innovation Centre, Glasgow G1 1RD, UK}

\author{Jonathan J.D. McKendry}
\affiliation{Institute of Photonics, Dept. of Physics, University of Strathclyde, Technology and Innovation Centre, Glasgow G1 1RD, UK}

\author{Gerald L. Leake}
\affiliation{State University of New York Polytechnic Institute, Albany, NY 12203, USA}

\author{Daniel J. Coleman}
\affiliation{State University of New York Polytechnic Institute, Albany, NY 12203, USA}

\author{Cung Tran}
\affiliation{State University of New York Polytechnic Institute, Albany, NY 12203, USA}

\author{Jeffrey St Louis}
\affiliation{State University of New York Polytechnic Institute, Albany, NY 12203, USA}

\author{John Mucci}
\affiliation{State University of New York Polytechnic Institute, Albany, NY 12203, USA}

\author{Cameron Horvath}
\affiliation{Applied Nanotools, Inc., Edmonton, AB T6G2M9, CA}

\author{Jocelyn N. Westwood-Bachman}
\affiliation{Applied Nanotools, Inc., Edmonton, AB T6G2M9, CA}

\author{Stefan F. Preble}
\affiliation{Microsystems Engineering, Rochester Institute of Technology, Rochester, NY 14623, USA}

\author{Martin D. Dawson}
\affiliation{Institute of Photonics, Dept. of Physics, University of Strathclyde, Technology and Innovation Centre, Glasgow G1 1RD, UK}

\author{Michael J. Strain}
\affiliation{Institute of Photonics, Dept. of Physics, University of Strathclyde, Technology and Innovation Centre, Glasgow G1 1RD, UK}

\author{Michael L. Fanto}
\affiliation{Air Force Research Laboratory, Information Directorate, Rome, NY, 13441, USA}

\author{Dirk R. Englund}
\email{englund@mit.edu}
\affiliation{Research Laboratory of Electronics, Massachusetts Institute of Technology, Cambridge, MA 02139, USA}

\begin{abstract}
Harnessing the full complexity of optical fields requires complete control of all degrees-of-freedom within a region of space and time --- an open goal for present-day spatial light modulators (SLMs), active metasurfaces, and optical phased arrays. Here, we solve this challenge with a programmable photonic crystal cavity array enabled by four key advances: (i) near-unity vertical coupling to high-finesse microcavities through inverse design, (ii) scalable fabrication by optimized, 300 mm full-wafer processing, (iii) picometer-precision resonance alignment using automated, closed-loop ``holographic trimming", and (iv) out-of-plane cavity control via a high-speed $\upmu$LED array. Combining each, we demonstrate near-complete spatiotemporal control of a 64-resonator, two-dimensional SLM with nanosecond- and femtojoule-order switching. Simultaneously operating wavelength-scale modes near the space- and time-bandwidth limits, this work opens a new regime of programmability at the fundamental limits of multimode optical control.
\end{abstract}

\maketitle

\section{Introduction}
\label{sec:intro}

Programmable optical transformations are of fundamental importance across science and engineering, from adaptive optics in astronomy \cite{Gardner2006} and neuroscience \cite{Packer2013,Marshel2019,Demas2021}, to dynamic matrix operations in machine learning accelerators \cite{Hamerly2019} and quantum computing \cite{Kok2007,Barredo2018,Ebadi2021}. Despite this importance, fast, energy-efficient, and compact manipulation of multimode optical systems --- the core objective of spatial light modulators (SLMs) --- remains an open challenge \cite{Shaltout2019a,Piccardo2021}. Specifically, the limited modulation bandwidth and/or pixel density of liquid crystal (LC) SLMs, digital micromirror displays, and other two-dimensional (2D) modulator arrays prevents complete control over the optical fields they tune.

\begin{figure*}
\includegraphics[width=\textwidth]{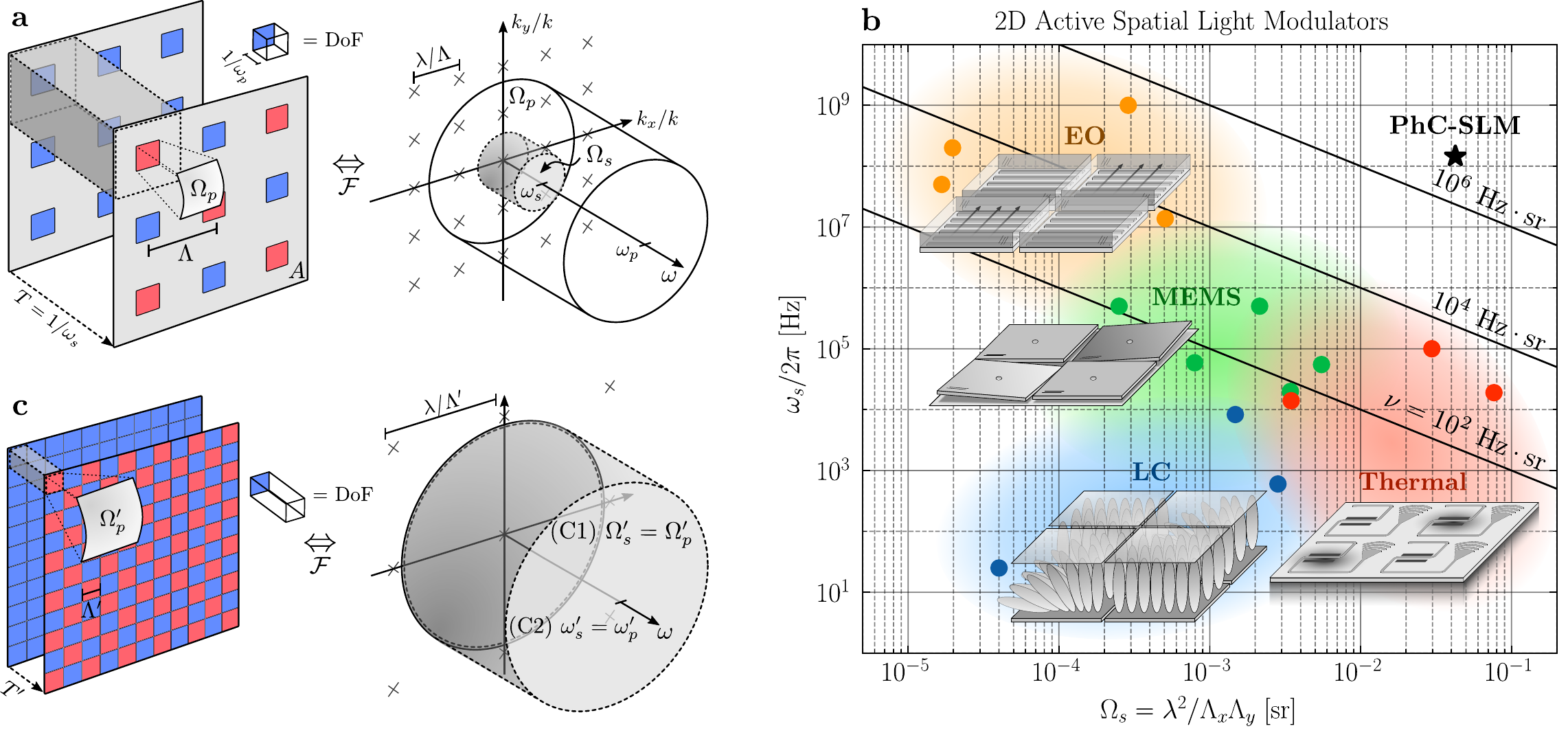}
\caption{\textbf{Full degree-of-freedom (DoF) spatiotemporal optical programming.} Present-day SLMs (a) feature a 2D array of $\Lambda$-pitch pixels within an aperture area $A$. Each pixel radiates at wavenumber $k = 2\pi/\lambda$ into the solid angle $\Omega_p$ and can be switched (blue $\leftrightarrow$ red color change indicates a $\pi$ phase change of the emitted field) over the timescale $T=1/\omega_s$ (given a modulation bandwidth $\omega_s$) with a large but slow fractional permittivity perturbation $\Delta\epsilon_p/\epsilon$ (e.g. liquid crystal rotation). The shaded volume indicates the smallest controllable near-field spatiotemporal mode. In the far-field (right), the corresponding shaded spatiotemporal bandwidth $\nu = \Omega_s\omega_s = (\lambda/\Lambda)^2\omega_s$ counts the controllable DoF per unit area and time in a single diffraction order. Trade-offs between $\omega_s$ and $\Omega_s$ in liquid crystal- (LC) \cite{McKnight1994, Marshel2019, Li2020}, thermal- \cite{Fatemi2019, Sun2013, Horie2017}, micro-electro-mechanical system- (MEMS) \cite{Shrauger2001, Yang2014, Wang2019, Bartlett2019, Zhang2022}, and electro-optic-driven (EO) \cite{Shuai2017, Junique2005, Smolyaninov2019, Benea2021} SLMs (b) limit $\nu \ll \Omega_p\omega_p$, the accessible pixel bandwidth given the delay-limited bandwidth $\omega_p \sim ck \Delta\epsilon_p/\epsilon$. Spatiotemporal control is thus limited and scattering into undesired diffraction orders (grey $\times$s) reduces diffraction efficiency. Alternatively, a fully-filled array of wavelength-scale resonant apertures (c) emitting into the solid angle $\Omega_p'$ can enhance the effect of fast (modulation frequency $\omega_s'$), low-energy perturbations $\Delta\epsilon_p' \ll \epsilon$ to simultaneously achieve space- and time-bandwidth limits (C1 and C2, respectively), yielding near-complete spatiotemporal control with $\nu' \approx \Omega_p'\omega_p'$.}
\centering
\label{fig:spatiotemporal}
\end{figure*}

Figure \ref{fig:spatiotemporal}a illustrates these limitations for a typical SLM comprised of a two-dimensional (2D), $\Lambda$-pitch array of tunable pixels (subscript $p$) emitting at wavelength $\lambda$ into the solid angle $\Omega_p$ with a system (subscript $s$) modulation bandwidth $\omega_s$. Given these parameters, each ``spatiotemporal" degree-of-freedom (DoF) simultaneously satisfying the minimum-uncertainty space- and time-bandwidth relations ($\delta A/\lambda^2 \cdot \delta\Omega =  1$ and $\delta t \cdot \delta \omega = 1$, respectively) can be illustrated as a real-space voxel with area $\lambda^2/\Omega_p$ and time duration $1/\omega_p$ for pixel bandwidth $\omega_p$. The optical-delay-limited pixel bandwidth $\omega_p\approx(\Delta\epsilon_p/\epsilon)ck$ can be approximated as a function of the achievable permittivity swing $\Delta\epsilon_p$ (for the speed of light $c$) using first-order perturbation theory \cite{Joannopoulos2008} or similarly derived from linear scattering theorems \cite{Miller2007}.

Integrating over the switching interval $T=1/\omega_s$ and aperture area $A$ then gives the total DoF count \cite{Gabor1961} \footnote{Note that the exact coefficient of proportionality in Eqn.~\ref{eqn:ff-dof} depends on the number of polarizations, the complex amplitude and phase controllability of each mode, and the exact definition of distinguishability when defining the Fourier uncertainty relations. For simplicity, we have omitted these $\mathcal{O}(1)$ coefficients.}
\begin{equation}
F = \int_{A,\Omega_p} \frac{\d A}{\lambda^2} \cdot \d\Omega \int_{T,\omega_p} \d t\cdot \d\omega.
\label{eqn:ff-dof}
\end{equation}
By comparison, the same switching period contains $N=A/\Lambda^2\leq F$ \textit{controllable} modes, each confined to the pixel area $\Lambda^2$ and time window $T$ (shaded box in Fig.~\ref{fig:spatiotemporal}a). Complete spatiotemporal control with $N=F$ is only achieved under the following criteria: (C1) emitters fully ``fill" the near-field aperture such that $\Omega_p$ matches the field-of-view $\Omega_s = (\lambda/\Lambda)^2$ of a single array diffraction order; and (C2) $\omega_s = \omega_p$. In the Fourier domain, the system's ``spatiotemporal bandwidth" $\nu=\Omega_s \omega_s$ counts the controllable DoF per unit area and time within a single far-field diffraction order. As illustrated by the shaded pillbox in  Fig.~\ref{fig:spatiotemporal}a, (C1) and (C2) are both satisfied when $\nu$ matches the accessible pixel bandwidth $\Omega_p\omega_p$.

Practical constraints have prevented present-day SLM technology from achieving this bound. In general, commercial devices approximate (C1) without achieving (C2). Specifically, they offer excellent near-field fill-factor across megapixel-scale apertures but use large $\mathcal{O}(\epsilon)$, slow index perturbations. Liquid crystal SLMs, for example, are limited to $\omega_s \sim 2\pi \times 10^3 \text{ Hz} \ll \omega_p$ by the slow rotation of viscous, anisotropic molecules that modulate the medium's phase delay \cite{Heilmeier1968,Zhang2014}. Digital micromirror-based SLMs offer moderately faster ($\resim 10^5$ Hz) binary amplitude modulation by displacing a mechanical reflector, but at the expense of diffraction efficiency \cite{Ren2015}. Mechanical phase shifters \cite{Hornbeck1983,Greenlee2011,Yang2014,Wang2019,Tzang2019} improve this efficiency but still require design trade-offs between pixel size and response time.

Recent research has focused on surmounting the speed limitations of commercial SLMs with integrated photonic phased arrays \cite{Sun2013, Chung2017, Poulton2019, Rogers2021} and active metasurfaces comprised of thermally \cite{Wang2016, Zhang2021, Wang2021}, mechanically \cite{Arbabi2018, Wang2019}, or electrically \cite{Smolyaninov2019, Wu2019, Park2021, Shirmanesh2020, Benea2021} actuated elements. These devices, however, do not satisfy (C1) (Table~\ref{table:slms}). Silicon photonics in particular has attracted significant interest due to its fabrication scalability; however, the combination of standard routing waveguides, high-power ($\resim$ mW/$\pi$ phase shift) thermal phase shifters, and vertical grating couplers in each pixel reduces the fill-factor of emitters, yielding $\Omega_p \gg \Omega_s$ \cite{Sun2013}. Scattering into the numerous diffraction orders within $\Omega_p$ then reduces the achievable zero-order and overall diffraction efficiencies ($\eta_0$ and $\eta$, respectively). For this reason, $\eta_0$ is a useful measure of near-field fill.  

Various workarounds, including 1D phased arrays with transverse wavelength tunability \cite{Chung2017,Kim2019a,Poulton2020}, sparse antenna arrays \cite{Fatemi2019}, and switched arrays \cite{Ito2020,Zhang2022} improve steering performance but restrict the spatiotemporal basis (i.e. limit $F$). Alternative nanophotonics-based approaches, often limited to 1D modulation, have their drawbacks as well: phase change materials \cite{Wang2016,Zhang2021,Wang2021} have slow crystallization rates and large switching energies, while electro-optic devices \cite{Huang2016, Shuai2017, Smolyaninov2019, Park2021, Shirmanesh2020, Benea2021, Ye2021}, to date, have primarily relied on large-area grating-based resonators to achieve appreciable modulation. 

\begin{figure}
\includegraphics[width=0.48\textwidth]{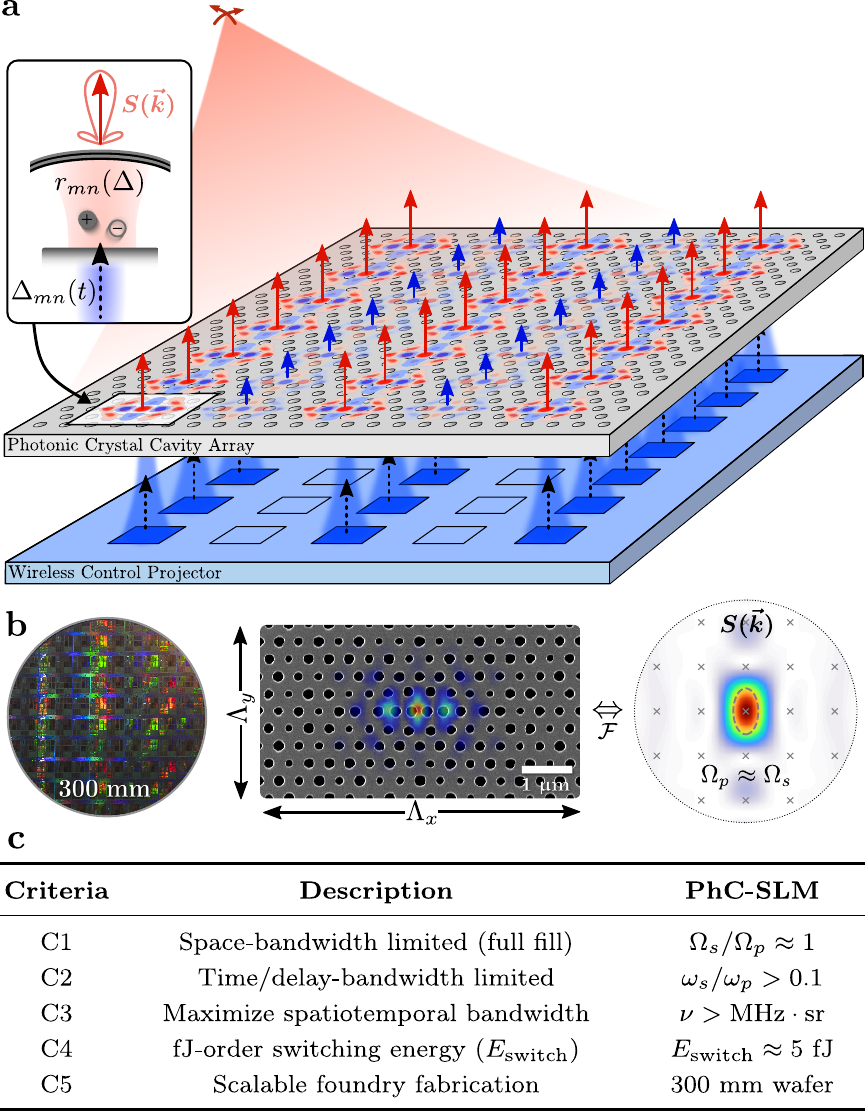}
\caption{\textbf{The photonic crystal spatial light modulator (PhC-SLM).} Complete spatiotemporal control is achieved by modulating an array of high-quality-factor ($Q>10^5$), small-mode-volume ($V < 0.1\lambda^3$) silicon PhC cavities with a high-speed incoherent $\upmu$LED array (a). Absorbed $\upmu$LED pulses control the detuning $\Delta$ of resonant pixels via free carrier dispersion, which varies the amplitude and phase (illustrated by the length and color, respectively, of emission arrows at each cell) of the pixel's complex reflection coefficient $r(\Delta)$. Despite sub-wavelength near-field confinement  (b, inset simulated mode profile overlaid on a SEM micrograph of an $L4/3$-type cavity \cite{Minkov2017}), each pixel is designed for directional ($\Omega_p\approx\Omega_s=\lambda^2/\Lambda_x\Lambda_y$) far-field scattering $S(\vec{k})$ into the zeroth diffraction order (marked by $\times$s) to satisfy (C1). Combining the reflection from each resonant ``antenna" in a large-scale aperture fabricated via optimized 300 mm wafer-scale processing (b, inset photograph) enables near-ideal SLM performance per the design criteria (C1-5) (c).}
\centering
\label{fig:overview}
\end{figure}

Figure \ref{fig:spatiotemporal}b compares the performance of these and other experimentally-demonstrated, active, 2D SLMs as a function spatiotemporal bandwidth's two components: modulation bandwidth $\omega_s$ and field-of-view $\Omega_s$. Controllability aside, the evident trade-off between these parameters illustrates the difficulty of creating fast, compact modulator arrays with high $\nu$. Thus, in addition to satisfying the complete control criteria (C1) and (C2), an ``ideal" SLM would (Fig.~\ref{fig:overview}c): (C3) maximize $\nu$ by combining wavelength-scale pitches (for full-field $\Omega_s\rightarrow 2\pi$ beamforming) with gigahertz (GHz)-order bandwidths $\omega_s$ competitive with electronic processors; (C4) support femtojoule (fJ)-order switching energies as desired for information processing applications \cite{Miller2017}; and (C5) have scalability to state-of-the-art megapixel-scale apertures.

These criteria motivate the resonant architecture in Fig.~\ref{fig:spatiotemporal}c. Here, (C3) and (C4) are achieved by switching a fully-filled array of wavelength-scale resonant optical antennas with fast, fJ-order perturbations $\Delta\epsilon_p/\epsilon\ll 1$. Each resonator's far-field scattering and quality factor $Q$ can then be tuned to achieve (C1) and (C2), respectively. \textit{Combined, this resonant SLM architecture enables complete, efficient control of the large spatiotemporal bandwidth supported by its constituent pixels. }

Figure~\ref{fig:overview} illustrates our specific implementation of this full-DoF resonant SLM: the photonic crystal spatial light modulator (PhC-SLM) \cite{Panuski2021}. Coherent signal light is reflected off a semiconductor slab (permittivity $\epsilon$) hosting a 2D array of semiconductor PhC cavities with instantaneous resonant frequency $\omega_0 + \Delta_{mn}(t)$. A short-wavelength incoherent control plane imaged onto the cavity array controls each resonator's detuning $\Delta_{mn}(t)\approx -\Delta \epsilon(t) / 2 \epsilon$ via the permittivity change $\Delta\epsilon_p(t)$ induced by photo-excited free carriers \cite{Soref1987, Panuski2019}. We optimize the resonator bandwidth $\Gamma\approx\omega_s\approx 2\pi\times \mathrm{GHz}$ (corresponding to a quality factor $Q = \omega_0/\Gamma \sim 10^5$) to maximize the linewidth-normalized detuning $\Delta/\Gamma$ without significantly attenuating the cavity's response at the carrier lifetime ($\tau$)-limited modulation rate $\omega_s = 1/\tau$. Under these conditions, free carrier dispersion efficiently modulates the complex cavity reflectivity $r(\Delta)$ to enable fast ($>100$ MHz given a $\resim$ns free carrier lifetime \cite{Tanabe2008}), low-energy (fJ-order) conversion of incoherent control light into a dense array of coherent, modulated signal modes (Appendix~\ref{appendix:analytic}). 

This out-of-plane, all-optical switching approach is motivated by the recent development of high-speed, high-brightness $\upmu$LED arrays \cite{Huang2020,Lin2020} integrated with complementary metal-oxide-semiconductor (CMOS) drive electronics for consumer displays \cite{Templier2018,Chen2019} and high-speed visible light communication \cite{Herrnsdorf2015,Ferreira2016}. In particular, gallium nitride $\upmu$LED arrays with GHz-order modulation bandwidths \cite{Ferreira2016, Cai2021}, sub-micron pixel pitches \cite{Park2021a}, and large pixel counts \cite{Hassan2021} have been demonstrated within the past few years. Applying these arrays for reconfigurable, ``wireless" all-optical cavity control eliminates electronic tuning elements at each pixel to avoid optical loss, pixel pitch limitations, and interconnect bottlenecks for planar architectures (as aperture area $A$ grows, $\mathcal{O}(A)$ pixel controls eventually cannot be routed through the $\mathcal{O}(\sqrt{A})$ perimeter) \cite{Miller2010}.

Free of these constraints, we designed high-finesse, vertically-coupled microcavities offering coupling efficiencies $>90\%$, phase-dominant reflection spectra \cite{Horie2017,Peng2019}, and directional emission $\Omega_p \approx \Omega_s$ for high-efficiency beamforming (Section~\ref{sec:qff}). Bespoke, wafer-scale processing allows us to fabricate these ``resonant antennas" in arrays with mean quality factors $\langle Q\rangle >10^6$ and sub-nm resonant wavelength standard deviation (Section~\ref{sec:foundryfab}). For fine tuning, we developed a parallel laser-assisted thermal oxidation \cite{Lee2009,Chen2011} protocol to then trim $8\times8$ cavity arrays to picometer-order uniformity \cite{Lee2009,Chen2011} (Section~\ref{sec:trimming}), enabling high-speed spatial light modulation with fJ-order switching energies and $\omega_s>2\pi \times 100$ MHz (Section~\ref{sec:switching}). Compared to the previous devices surveyed in Fig.~\ref{fig:spatiotemporal}b, our PhC-SLM offers near-complete control over an order-of-magnitude larger spatiotemporal bandwidth. 

\section{Inverse-Designed Resonant Pixels}
\label{sec:qff}
\begin{figure}
\includegraphics[width=0.5\textwidth]{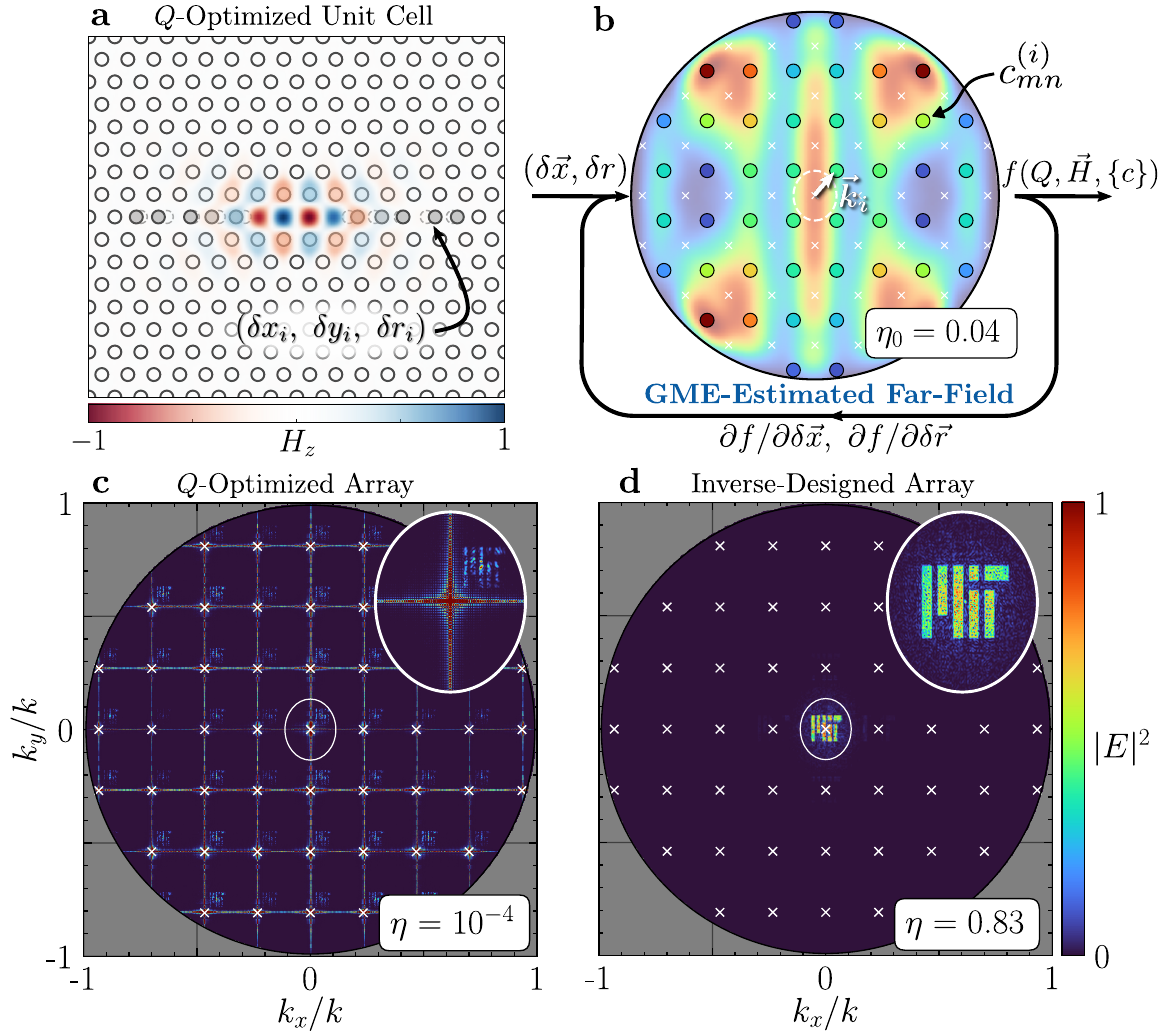}
\caption{\textbf{Optimized holography with inverse-designed, vertically-coupled microcavity arrays.} (a) Silicon $L3$ slab defect cavity design (hexagonal lattice constant $a = 0.4~\upmu$m; hole radius $r/a=0.25$; slab thickness $t=220$ nm) with overlaid midplane magnetic field profile $H_z$ after $Q$ optimization by displacing ($\delta x_i,~\delta y_i$) and resizing ($\delta r_i$) the shaded holes in the $16a\times16(\sqrt{3}/2)a$ periodic unit cell. Hole shifts are magnified by $3\times$ for visualization. The confined cavity mode radiates into the broad far-field profile in (b, background), violating (C1) and yielding a zero-order diffraction efficiency $\eta_0\ll 1$. As a result, simulated trial holograms (c) from a $64\times 64$ cavity array with optimized detunings (Appendix~\ref{appendix:convex}) have minimal overall diffraction efficiency $\eta$. Inverse design (b) solves these problems. Guided mode expansion (GME) approximates the mode's $Q$ \textit{and} far-field profile by sampling the losses $\lbrace c \rbrace$ at the array's diffraction orders (white $\times$s) displaced by Bloch boundary conditions $\vec{k}_i$ (i.e. at the colored dots). An objective function $f$ that maximizes $Q$, confines $\vec{H}$, and minimizes $\lbrace c \rbrace$ at any non-zero diffraction order can then be efficiently optimized with respect to \textit{all} hole parameters using reverse-mode automatic differentiation (b). The resulting devices with high-$Q$, efficient coupling, and directional emission enable high-performance ($\eta \sim 1$) resonant holography (d).}
\centering
\label{fig:holograms}
\end{figure}

\begin{figure*}
\includegraphics[width=\textwidth]{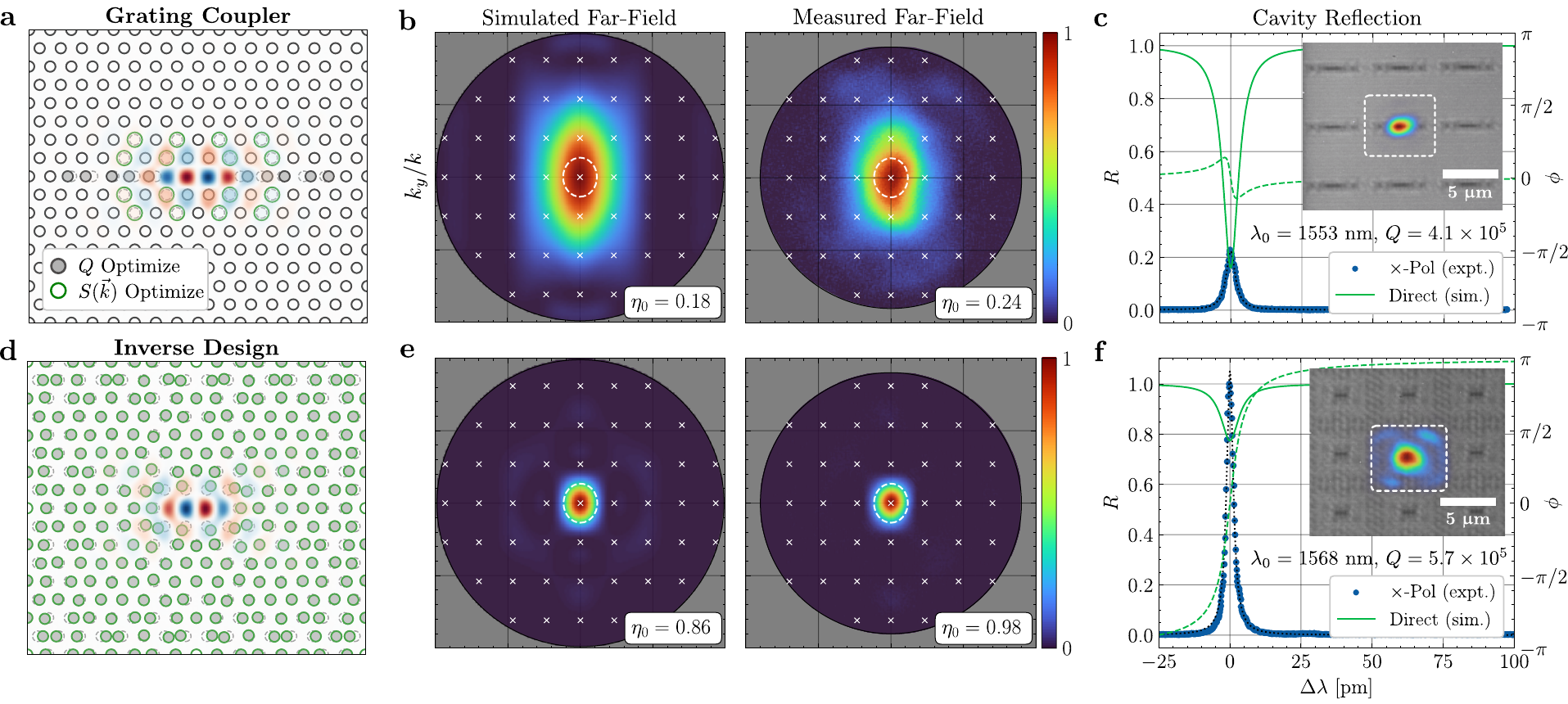}
\caption{\textbf{Experimental comparison of existing (a-c) and inverse-designed PhC cavities with high-$Q$ and near-diffraction-limited vertical beaming.} Superimposing a grating perturbation (a, green; $\delta r_i$ magnified by $20\times$ for visualization) on the $Q$-optimized design of Fig.~\ref{fig:holograms}a improves vertical coupling at the expense of reduced $Q$, yielding the simulated far-field intensity profile in (b, left) with $\eta_0=0.18$. Our measured far-field profile (b, right), collected from a grating-coupled cavity using a cross-polarized imaging setup (Appendix~\ref{appendix:setup}), confirms the broad emission relative to the array field-of-view (dashed white line) $\Omega_s$. This mismatch explains the low effective ``fill factor" and poor coupling observed in our resonant imaging (c, inset) and near-field reflection spectra (c, blue), respectively. An input Gaussian beam (with waist matched to the unit cell dimensions) is undercoupled and exhibits an amplitude-dominant power reflectivity $R=|r|^2$ modulation (c, solid green) with low phase variation $\Delta \phi$ (c, dashed green). Our inverse designed cavities (d) overcome these issues by optimizing \textit{every} hole in the unit cell to vertically scatter cavity leakage for any target $Q$, producing ``ideal" resonant SLM pixels satisfying (C1). Specifically, they support near-diffraction-limited emission (e) with $\eta_0\sim1$ due to fully-filled near-field resonant scattering (f, inset), a $\resim 5 \times$ experimental resonance contrast enhancement (f), and $>94\%$ single-sided (i.e. assuming an ideal back-reflector described in Appendix~\ref{appendix:substrate}) coupling to an input Gaussian beam for phase-dominant modulation (f, green).}
\centering
\label{fig:singlecav}
\end{figure*}

The sub-wavelength (i.e. normalized volume $\tilde{V} = V/(\lambda/n)^3 < 1$ relative to a cubic wavelength in the confining dielectric of refractive index $n$), high-$Q$ (up to $\resim 10^7$) \cite{Asano2017,Hu2018} modes of 2D PhC cavities enable (C4) \cite{Nozaki2010}, but at the expense of (C1) since $Q$ optimization (via hole displacements as in Fig.~\ref{fig:holograms}a \cite{Minkov2014}) cancels radiative leakage. The exact displacement parameters are typically numerically optimized in computationally expensive finite-difference time-domain (FDTD) simulations, which ultimately limits the number of free parameters. Compared to the ideal apertures in Fig.~\ref{fig:spatiotemporal}c, the optimized cavity unit cell confines a spatially complex mode with $\Omega_p \gg \Omega_s$ (Fig.~\ref{fig:singlecav}b, background), violating (C1) and limiting the zero-order diffraction efficiency to $\eta_0\approx 0.04$. The result is poor beamforming performance as exemplified by the distorted, low-efficiency far-field pattern emitted by a $64\times 64$ cavity array with optimized detunings (derived with the algorithm in Appendix~\ref{appendix:convex}) to match a target far-field image (MIT logo).
 
Fortunately, these limitations are not fundamental: the effective scattering aperture $A_0 = \lambda^2/\Omega_p$ of a resonant mode can extend beyond its $1/e$ decay area $A_e$. This apparent space-bandwidth violation $(A_e/\lambda^2)\cdot \Omega_p = A_e/A_0 < 1$ is enabled by resonant scattering from the mode's evanescent field, which raises the basic question: how should scatterers be arranged to produce a desired far-field emission pattern?

One established approach is a harmonic $2a$-period grating perturbation (Fig.~\ref{fig:singlecav}a) that ``folds" energy concentrated at the band-edge $k_x = \pi/a$ back to $k_{||}=0$, yielding vertical radiation at the expense of reduced $Q$ \cite{Tran2009,Tran2010,Portalupi2010,Qiu2012}. In the perturbative regime, the far-field scattering profile is an image of the broad band-edge mode. Thus, once the grating-induced loss becomes dominant, further magnifying the perturbation reduces $Q$ without significantly improving directivity. Fig.~\ref{fig:singlecav}b shows the narrowed far-field profile produced by a $\delta r_i/r\approx 0.02$ grating perturbation, which balances the reduced $Q\approx 8\times 10^5$ and a modest diffraction efficiency improvement ($\eta_0=0.18$).

By contrast, our design strategy (Fig.~\ref{fig:holograms}b) combines semi-analytic guided mode expansion (GME) simulations with automatic differentiation to maximize $\eta_0$ (and thereby the effective near-field fill factor) for a given target $Q$ using \textit{all} of the hole parameters. In each iteration, GME approximates the cavity eigenmode and radiative loss rates $c_{mn}^{(i)}$ at each of the array's reciprocal lattice vectors  (i.e. diffraction orders) offset by the Bloch periodic boundary conditions $\vec{k}_i$ \cite{Andreani2006}. These coupling coefficients coarsely sample the cavity's approximate far-field emission (Appendix~\ref{appendix:gme}). Scanning $\vec{k}_i$ over the irreducible Brillouin zone of the rectangular cavity array improves the sampling resolution, and an overall $Q$ can be estimated by averaging the total loss rates $\Gamma^{(i)} = \sum_{mn} c_{mn}^{(i)}$ in each simulation. Reverse-mode automatic differentiation then allows us to efficiently optimize an objective function
\begin{equation}
    f = \frac{1}{N}\sum_{i=1}^N \frac{c_{00}^{(i)}}{\Gamma^{(i)}} \arctan \left(\frac{Q}{Q_0} \right)|E_0|^2
    \label{eqn:objective}
\end{equation}
targeting three main goals: 1) increase $Q$ to a design value $Q_0$; 2) force the associated radiative loss into the array's zeroth diffraction order for efficient vertical coupling; and 3) minimize $V$ by maximizing $|E_0|$, the electric-field magnitude at the center of the unit cell. The resulting designs support tunable-$Q$ resonances with near-diffraction-limited ($\Omega_p \approx \Omega_p$) vertical beaming comparable to the ideal planar apertures of Fig.~\ref{fig:spatiotemporal}c. The example design of Fig.~\ref{fig:singlecav}d, for instance, maintains $Q\approx 8\times 10^5$ with $\eta_0=0.86$ based on the simulated far-field profile in Fig.~\ref{fig:singlecav}e.

We prototyped each design at a commercial electron beam lithography (EBL) foundry \footnote{Applied Nanotools, Inc. \url{https://www.appliednt.com/}} before transitioning to the wafer-scale foundry process described in Sec.~\ref{sec:foundryfab}. The near- and far-field reflection characteristics of the fabricated devices were measured with the cross-polarized microscopy setup detailed in Appendix \ref{appendix:setup}. Fig.~\ref{fig:singlecav}b-c and Fig.~\ref{fig:singlecav}e-f show the results for the grating coupled and inverse-designed cavities, respectively. The optimal grating-coupled cavities offer $Q\sim 4\times 10^5$ at $\lambda \approx 1553 \text{ nm}$ with a near-field resonant scattering profile well-centered on the cavity defect (Fig.~\ref{fig:singlecav}c, inset). The mode mismatch between this wavelength-scale PhC mode and the wide-field input beam (Gaussian beam with $\resim 150~ \upmu\text{m}$ waist diameter for array-level excitation) is further evidenced by the small normalized reflection amplitude (relative to that of the inverse designed cavities) on resonance as well as the broad far-field profile (Fig.~\ref{fig:singlecav}b) with $\eta_0=0.24$.

By comparison, inverse design non-perturbatively modifies the cavity mode (Fig.~\ref{fig:singlecav}d) to produce the near-ideal measured far-field profile in Fig.~\ref{fig:singlecav}e satisfying (C1) with $\eta_0 = 0.98$ while \textit{simultaneously} increasing $Q$ to $5.7\times 10^5$. We attribute the slight increase in zero-order diffraction efficiency over the simulated value $\eta_0 = 0.86$ to the substrate-dependent effects described in Appendix~\ref{appendix:substrate}. The fully-filled near-field resonant scattering image in Fig.~\ref{fig:singlecav}f explains the close resemblance between this measured $S(\vec{k})$ and that of an ideal uniform aperture \cite{Hansen1981}. In addition, the narrowed emission profile $S(\vec{k})$ yields a $\resim 5 \times$ increase in cross-polarized reflection and the phase-dominant simulated direct reflection spectrum in Fig.~\ref{fig:singlecav}f. The latter is achieved by 94\% one-sided coupling to a Gaussian beam with optimized waist diameter (Appendix \ref{appendix:opt-overlap}).

Combined, these results break the traditional coupling--$Q$ tradeoff (offering an order-of-magnitude improvement in the figure-of-merit $\eta_0 \cdot Q$ for the prototype devices in Fig.~\ref{fig:singlecav}) to enable high-performance beamforming at the space-bandwidth limit (C1). These results are supported by the simulated hologram in Fig.~\ref{fig:holograms}d: an array of optimally detuned, inverse-designed cavities forms a clear far-field image with a several order-of-magnitude improvement in overall diffraction efficiency ($\eta=0.83$) over existing designs.

\section{Foundry-Fabricated High-Finesse Microcavity Arrays}
\label{sec:foundryfab}

While EBL enables fabrication of few-pixel prototypes with state-of-the-art resolution and accuracy, serial direct-write techniques do not satisfy (C5). Field stitching issues and sample preparation aside, a single cm$^2$, megapixel-scale sample would require a full day of EBL write time alone. We therefore developed a full-wafer deep-ultraviolet photolithography process specifically optimized for wavelength-pitch arrays of high-$Q/V$ PhC microcavities in a commercial foundry \cite{Fahrenkopf2019}. 

\begin{figure}
\includegraphics[width=0.42\textwidth]{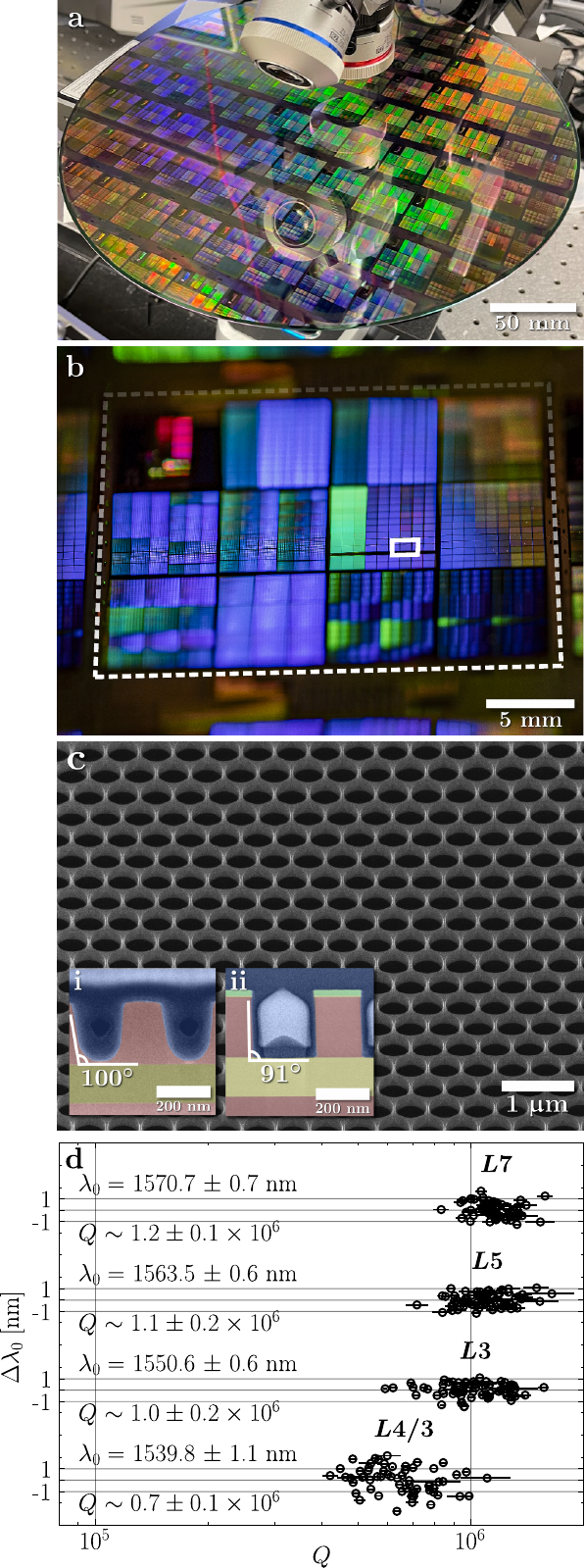}
\caption{\textbf{Full-wafer photonic crystal fabrication in an optimized 300 mm foundry process.} A wafer (a) contains 64 complete reticles (b) each comprising millions of inverse designed PhC cavities. The before (i) and after (ii) false-color (blue: metal fill; red: silicon; yellow: silicon dioxide; green: etch mask) transmission electron microscope cross-sections show how process optimization enables high-quality PhC lattices (c) that support $Lm$-type cavity arrays with $\langle Q \rangle > 10^6$ and sub-nanometer wavelength standard deviation (d).}
\centering
\label{fig:foundryfab}
\end{figure}

A central goal was to create vertical etch side-walls. The transmission electron microscope (TEM) cross-section in Fig.~\ref{fig:foundryfab}i shows that the default fabrication process (optimized for isolated waveguides) yielded an oblique ($100^\circ$), incomplete etch through the silicon device layer for the target PhC lattice parameters. Both nonidealities erase the membrane's vertical reflection symmetry, leading to coupling between even- and odd- symmetry (about the slab midplane) modes that ultimately limits the achievable $Q$ of bandgap-confined resonances \cite{Asano2006}. By contrast, our revised fabrication process achieves near-vertical $91^\circ$ sidewall angles (Fig.~\ref{fig:foundryfab}ii), yielding  high-quality PhC lattices for a range of hole diameters between the $\resim 100$ nm critical-dimension  and $2r\approx a$ (Fig.~\ref{fig:foundryfab}c). Using TEM cross-sectioning and automated optical metrology as feedback over multiple 300 mm wafer runs in the \textit{AIM Photonics} foundry's 193 nm DUV water-immersion lithography line, this new process relies on a combination of dose-optimized reverse (positive) tone lithography, high-accuracy laser written masks, and optimized etch termination. Following fabrication and dicing, we post-processed individual die with a backside silicon nitride anti-reflection coating and, as required, suspend the PhC membranes with a timed wet etch.

The resulting die contain isolated and arrayed PhC cavities with swept dimensions to offset systematic fabrication biases. We chose $Lm$-type cavity designs --- formed by removing $m$ holes from the PhC lattice as demonstrated by the $L3$ unit cells in Fig.~\ref{fig:singlecav} --- to host tunable-volume (via variable $m$), high-$Q$ resonant modes with even reflection symmetry (about the unit cell axes) as required for vertical emission \cite{Kim2006}. The highest-performance isolated devices feature $Q>10^6$ with normalized volumes $\tilde{V} \approx 0.3$. With a joint spectral- and spatial-confinement (quantified by the figure-of-merit $Q/\tilde{V}) \approx 4\times 10^6$, these devices are among the highest-finesse optical cavities ever fabricated in a foundry process. 

Our optimized foundry processing extends this exceptional single-device performance (rivaling record EBL-fabricated devices) to large-scale cavity arrays. We developed a fully-automated measurement system (Appendix~\ref{appendix:setup}) to locate and characterize hundreds of cavities per second via parallel camera readout. The resulting data, extracted from over $10^5$ devices measured across the wafer, allow us to statistically analyze resonator performance and fabrication variability at the die, reticle, and wafer level. Fig.~\ref{fig:foundryfab}d, for example, shows resonant wavelength and $Q$ variations within $8\times 8$ arrays of four different cavity designs. Using camera readout of the reflected wide-field excitation, each data set is extracted from a single wavelength scan of a tunable laser. Besides the expected correlation between uniformity and mode volume \cite{Sekoguchi2014}, the data demonstrates --- for the first time, to our knowledge --- the ability to fabricate sub-wavelength ($\tilde{V}<1$) microcavity arrays with $\langle Q \rangle > 10^6$ and sub-nanometer resonant wavelength standard deviation ($\sigma_\lambda\approx 0.6$~nm).  Critically for beamforming, this uniformity also extends to the far-field: Appendix~\ref{appendix:uniformity} shows that each cavity in an $8\times 8$ array emits vertically with $\eta_0 = 0.86 \pm 0.07$, in quantitative agreement with the simulated result in Fig.~\ref{fig:singlecav}. 

\section{Holographic Trimming}
\label{sec:trimming}

\begin{figure*}
\includegraphics[width=\textwidth]{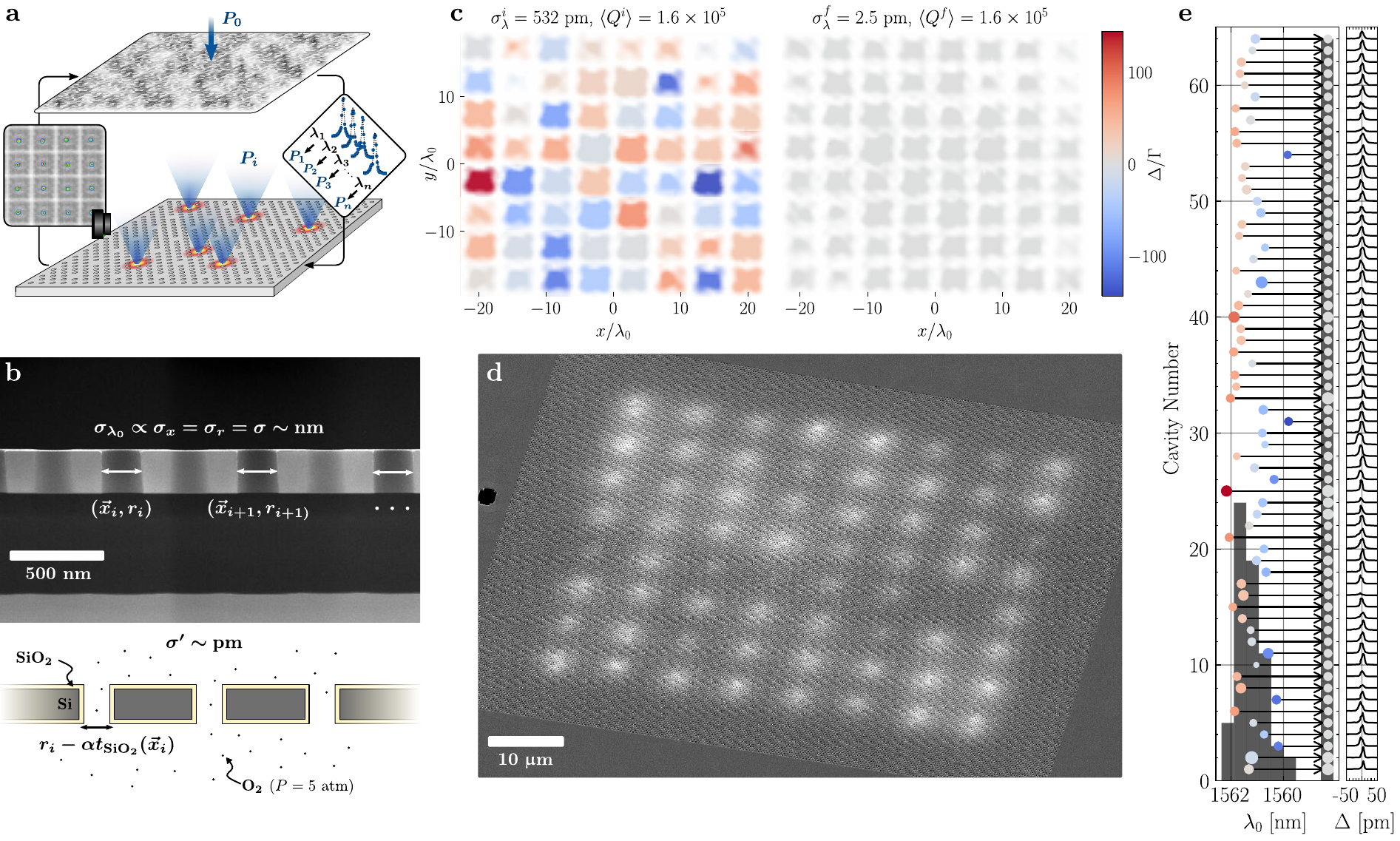}
\caption{\textbf{Parallel, fully-automated, and low-loss microcavity trimming via structured laser oxidation.} In each iteration of the trimming loop (a), a weighted GS algorithm distributes a visible trimming laser with power $P_0$ to powers $\lbrace P_i \rbrace$ at desired cavities based on the measured resonant wavelength $\lambda_i$ of each cavity. A few nm-thick layer of thermal oxide grows at each optical focus (photographed spots in the inset cavity array image), reducing the as-fabricated standard deviation in hole parameters $\sigma\sim \text{nm}$ (b) and permanently shifting the targeted resonances. The initial and final near-field hyperspectral reflection images (c, color-coded by each device's wavelength normalized detuning $\Delta/\Gamma$) show the $>100\times$ reduction in resonant-wavelength standard deviation to $\sigma_\lambda = 2.5~\mathrm{pm}$ without affecting the mean quality factor $Q > 10^5$. The effective dimensions of the final array (d) are thus homogenized to $\sigma'\sim \text{pm}$ length-scales by oxidation. Regions of local oxide growth in helium ion microscopy images of the trimmed PhC-SLM (d) appear as bright areas. The device summary (e) shows the wavelength shift, $Q$ variation (quantified by dot area), and aligned reflection spectra of each device.}
\centering
\label{fig:trimming}
\end{figure*}

In addition to these overlapping far-field emission profiles, programmable multimode interference requires each cavity to operate near a common resonant wavelength $\lambda_0$. For sufficiently high-$Q$ resonators, this tolerance cannot be solely achieved through optimized fabrication since $\mathcal{O}(\text{nm})$ fabrication fluctuations translate to $\mathcal{O}(\text{nm})$ resonant wavelength variations \cite{Taguchi2011,Minkov2013}. Our prototype $8\times 8$ arrays of $L3$ cavities (chosen to optimally balance requirements on $Q$, $V$, directive emission, and fabrication tolerance) typically span a $\resim 3~\text{nm}$ peak-to-peak wavelength variation (given $\sigma_\lambda\approx 0.6$ nm), corresponding to hundreds of linewidths for the target $Q\sim 10^5$. 

To correct this nonuniformity, we developed an automated, low-loss, and picometer-precision trimming procedure based on laser-assisted thermal oxidation (Fig.~\ref{fig:trimming}). Two features of our approach resolve the speed and controllability limitations of prior single-device implementations \cite{Lee2009,Chen2011}: 1) accelerated oxidation in a high-pressure chamber with in-situ characterization; and 2) holographic fanout of the trimming laser to simultaneously address multiple devices. In each iteration of the automated trimming loop (Fig.~\ref{fig:trimming}a), the resonant wavelengths $\lbrace \lambda_i\rbrace$ are measured and a subset $T$ containing $N$ devices is selected to maximize the total trimming distance $N(\min_T \lbrace \lambda_i \rbrace  - \lambda_t)$ to a target wavelength $\lambda_t$. Each cavity in $T$ is then targeted by a visible laser distributed by the liquid crystal SLM setup described in Appendix~\ref{appendix:setup}. To generate the required phase masks, we developed an open-source, GPU-accelerated experimental holography software package that implements fixed-phase, weighted Gerchberg-Saxton (GS) phase retrieval algorithms \footnote{slm-suite. \url{https://github.mit.edu/cpanuski/qp-slm}}. Using camera feedback, the algorithm can generate thousands of near-diffraction-limited foci with $\resim 1\%$ peak-to-peak power uniformity and single-camera-pixel-order location accuracy within a few iterations (Appendix~\ref{appendix:qpslm}).

The holographically-targeted pixels are then laser-heated with a computed exposure power and duration (based on the current trimming rates, resonance locations, and other array characteristics) to grow thermal oxide at the membrane surface. For thin oxide layers, the consumption of silicon during the reaction with ambient oxygen permanently blueshifts the cavity resonance in proportion to the oxide thickness $t_{\text{SiO}_2}$ (Fig.~\ref{fig:trimming}b) \cite{Chen2011}. Per the Deal-Grove model, the rate-limiting diffusion of oxygen through the grown oxide accelerates with increasing oxygen pressure --- a well-known technique in microelectronics fabrication \cite{Lie1982}. We therefore oxidize our samples in pure oxygen with partial pressure $P_{\text{O}_2} = 5~\text{bar}$, enabling $\d\lambda_0/\d t \approx 0.1 ~\text{nm/s}$ resonance trimming rates over $\Delta\lambda_0>20~\text{nm}$ wavelength ranges. After each trimming exposure, we remeasure the resonance statistics and recycle the loop until all devices are aligned within a set tolerance about $\lambda_t$. The trimming algorithm also accounts for long-term moisture adsorption to the membrane surface, thermal cross-talk, and trimming rate variations (Appendix~\ref{appendix:trimming}). 

Fig.~\ref{fig:trimming} demonstrates the results of this trimming procedure applied to our prototype $8\times 8$ pixel PhC-SLM. Prior to trimming, the hyperspectral near-field reflection image in Fig.~\ref{fig:trimming}c shows the large ($>200$ linewidths for the mean quality factor $\langle Q \rangle =  1.6\times 10^5$) resonant wavelength variation between the otherwise spatially uniform and high-fill resonant modes. Holographic trimming reduces the wavelength standard deviation and peak-to-peak spread by $>100\times$ to $\sigma_\lambda=2.5~\text{pm}$ and $\Delta \lambda_0^\text{p-p} = 1.3\Gamma = 13~\text{pm}$, respectively, enabling all 64 devices (imaged in Fig.~\ref{fig:trimming}d) to be resonantly excited at a common operating wavelength (Fig.~\ref{fig:trimming}e). Since $\sigma_\lambda$ is directly related to the corresponding hole radius and placement variability ($\sigma_r$ and $\sigma_h$, respectively) with an $\mathcal{O}(1)$ design-dependent constant of proportionality, the thermal oxide homogenizes the effective dimensions of each microcavity to the picometer scale. The mean quality factor and near-field reflection profile of the array remain largely unmodified throughout the process as evidenced by Fig.~\ref{fig:trimming}c and Fig.~\ref{fig:trimming}e.

To our knowledge, these results are the first demonstration of parallel, in situ, non-volatile microcavity trimming. The achievable scale is currently limited by environmental factors that could be overcome with stricter process control (Appendix~\ref{appendix:trimming}). Even without these improvements, the current uniformity, scale, and induced loss outperform the corresponding metrics of the previous techniques reviewed in Appendix~\ref{appendix:trimming}, paving the way towards scalable integrated photonics with high-$Q$ resonators. 

\section{All-Optical Spatial Light Modulation}
\label{sec:switching}

Once trimmed to within a linewidth, each resonator reflects an incident coherent field $E_\text{i}(\vec{r},t)$, producing a far-field output \cite{Haus1984}
\begin{equation}
    E_\text{r}(\vec{k},t) = S(\vec{k})\sum_{m,n} r\lbrace \Delta_{mn}(t)\rbrace E_\text{i}(\vec{r}_{mn},t) e^{j\vec{k} \cdot \vec{r}_{mn}}
    \label{eqn:far-field}
\end{equation}
that can be dynamically controlled within $S(\vec{k})$ by setting the detuning $\Delta_{mn}(t)$ (and therefore the near-field reflection coefficient $r$) of each resonator. Experimentally, we measure the intensity pattern $|E_r(\vec{k})|^2$ on the back focal plane of a microscope objective above the PhC-SLM (as with the single-device characterization in Sec.~\ref{sec:qff}) and optically program $\Delta_{mn}(t)$ via photo-excited free carriers. The corresponding setups are detailed in Appendix~\ref{appendix:setup}. 

\begin{figure*}
\includegraphics[width=\textwidth]{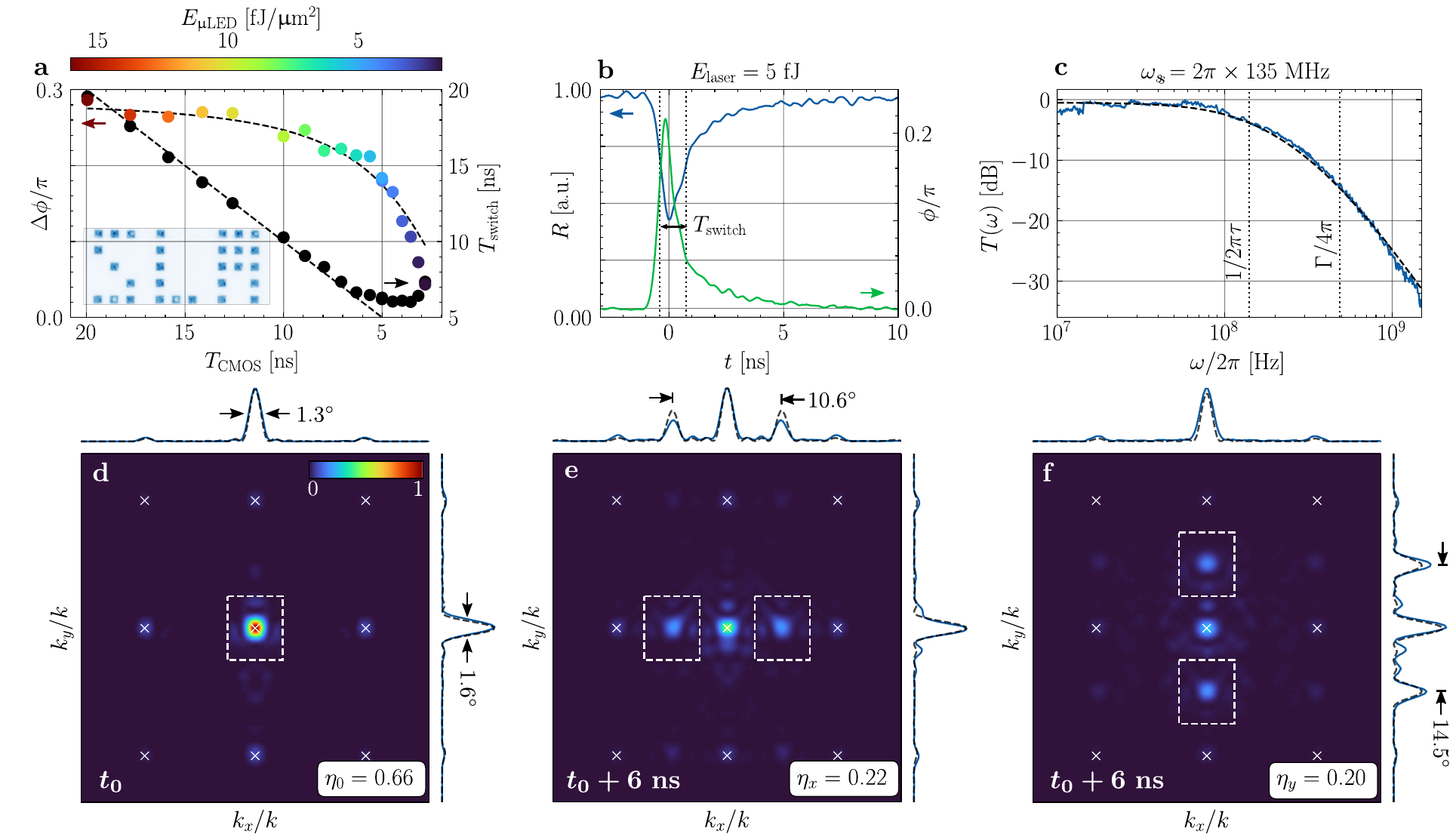}
\caption{\textbf{Nanosecond switching (a-c) and spatial light modulation (d-f).} (a) Peak phase shift $\Delta\phi$ and half-maximum switching interval $T_\text{switch}$ produced by pulses from a CMOS-integrated $\upmu$LED array imaged onto the cavity array (inset: cavities illuminated to form the letters `S', `L', and `M') as a function of trigger duration $T_\text{CMOS}$ and pulse energy density $E_{\upmu\text{LED}}$. (b) Complex reflectivity ($r = \sqrt{R}e^{j\phi}$) modulation with femtojoule-order pulse energies $E_\text{laser}$ from a focused visible laser. (c) Output probe to input visible (pump) power transfer function $T(\omega)$ fit to a second-order response function, yielding $\omega_s=2\pi \times 135$ MHz limited by the free carrier lifetime $\tau\approx 1.1$ ns and cavity bandwidth $\Gamma \approx 1$ GHz. (d) Far-field intensity profile, half-maximum beam widths, and zero-order diffraction efficiency $\eta_0$ (integrated within the dashed white box) of the trimmed array at time $t_0$ with horizontal and vertical cross-sectional profiles (blue traces) compared to those of an $8\times 8$ array of planar apertures with 80\% linear fill (black, dashed). (e/f) Analogous results for the switched array with an optically-patterned horizontal (vertical) amplitude grating at the maximum extinction time $t_0+6$ ns, producing $\pm 1^\text{st}$-order diffraction peaks over a $10.6^\circ$ ($14.5^\circ$) field-of-view and diffraction efficiency $\eta_x = 0.22$ ($\eta_y = 0.20$).}
\centering
\label{fig:switching}
\end{figure*}

Absent a control input ($\Delta_{mn} \approx 0$), Fig.~\ref{fig:switching}d shows the static far-field intensity pattern $|E_r(\vec{k})|^2$ of a wide-field-illuminated (i.e. $E_i(\vec{r}) \approx E_i$) $8\times 8$ trimmed array with $\lbrace Q \rbrace = 1.85 \times 10^5$ and $\sigma_\lambda = 5$ pm at $\lambda = 1562$ nm. The inverse-designed cavity unit cells minimize scattering into undesired diffraction orders, producing a high-efficiency ($\eta_0=0.66$) zero-order beam with the expected $1.3^\circ$ and $1.6^\circ$ horizontal and vertical beamwidths given the $42.0\lambda\times 36.4\lambda$ aperture size. The cross-sectional beam profiles are well matched to the simulated emission profile of uniform apertures with width $w=0.8\lambda$, suggesting an 80\% effective linear fill of the array. This extracted value agrees with the observed zero-order efficiency and the array's physical design (each $16a\times 16a$ cavity offering near-unity fill was padded to $20a\times 20a$ to limit coupling to adjacent cells).

After confirming the static performance of the array, we conducted optical switching experiments with two sources: an incoherent $\upmu$LED array and a pulsed visible laser. The $\upmu$LED array contains $16\times 16$ individually-addressable gallium nitride $\upmu$LEDs with $>$150 MHz small-signal bandwidth and $\resim 10^6$ cd/m$^2$ peak luminances (at 450 nm) flip-chip bonded to high-efficiency CMOS drivers \cite{Zhang2013, Herrnsdorf2015}. Using the setup in Appendix~\ref{appendix:setup}, we imaged the 100 $\upmu$m-pitch array with variable demagnification and rotation onto the PhC cavity array. Digitally triggering the CMOS drivers then enables reconfigurable, binary optical addressing as illustrated by the imaged projections of three letters on the PhC-SLM (Fig.~\ref{fig:switching}a). We measured the resulting pixel reflection amplitude and phase using locked, shot-noise-limited balanced homodyne detection (Appendix~\ref{appendix:setup}). Fig.~\ref{fig:switching}a depicts the maximum phase shift $\Delta\phi$ as a function of CMOS trigger duration $T_\text{CMOS}$ and imaged pump energy density $E_{\upmu\text{LED}}$. Single-cavity switching is possible with energy densities below $10$ fJ/$\upmu$m$^2$ (corresponding to $\resim 100$ fJ total energy for our chosen demagnification) and a minimum trigger duration $T_\text{CMOS} \approx 5$ ns. Shorter trigger pulses produce relatively constant-width pulses (due to the $\upmu$LED fall time) with insufficient energy for high-contrast switching. 

Confining visible pump pulses in space and time to the silicon free-carrier diffusion length ($\sim 1~\upmu$m) and lifetime ($\tau \approx 1$~ns), respectively, would reduce the required switching energy and maximize bandwidth bandwidth. While either metric is achievable with existing $\upmu$LED arrays \cite{McKendry2009,Park2021a} and optimization to achieve both simultaneously is ongoing \cite{Lan2020}, we demonstrated the expected performance enhancement with a pulsed visible ($\lambda = 515$ nm) laser. Fig.~\ref{fig:switching}b shows that 3 dB power reflectivity changes and high-contrast phase modulation are feasible for 5 fJ pump pulses over a switching interval $T_\text{switch}\approx 1$ ns, thereby satisfying (C4). Free-carrier dispersion is the dominant switching mechanism for these isolated, ns-order switching events (Appendix~\ref{appendix:analytic}). While repeated switching over $\upmu$s-order timescales leads to a slowly-varying thermo-optic detuning \cite{Barclay2005}, various optical communications techniques (constant-duty line codes, for example \cite{Winzer2006}) can maintain average device temperature during high-speed free carrier modulation. To demonstrate this decoupling of switching mechanisms, we measured the normalized small-signal transfer function $T(\omega)$ between a harmonic pump power (produced by a network analyzer-driven amplitude electro-optic modulator) and the phase-locked homodyne response. When aligned to the thermally-detuned resonance, the results (Fig.~\ref{fig:switching}c) match the expected second-order response  $T(\omega)=1/\lbrace \left[ 1+(\omega \tau)^2 \right] \left[ 1+(\omega/\pi\Gamma)^2 \right] \rbrace$ set by carrier- and cavity-lifetime-limited bandwidths ($1/\tau$ for a fitted carrier lifetime $\tau = 1.1$~ns and the measured $\Gamma= 1.0$~GHz, respectively). While satisfying (C2) therefore requires higher-$Q$ resonators, the current regime of operation enables near-complete control over a larger bandwidth $\omega_s=2\pi\times 135~\text{MHz}\approx 1/\tau$, i.e. without significantly degrading the carrier-lifetime-limited modulation bandwidth.

Combining this optimized switching with the space-bandwidth-limited vertical beaming of each resonator enables multimode programmable optics approaching the fundamental limits of spatiotemporal control. We currently probe the PhC-SLM in a wide-field, cross-polarized setup that produces amplitude-dominant Lorentzian reflection profiles $r(\Delta) \propto 1/(1+j\Delta)$ regardless of the resonator coupling regime (cavity emission is isolated from specular reflection). For simplicity, we therefore conducted proof-of-concept demonstrations using the PhC-SLM as an array of high-speed binary amplitude modulators. In this modality, a nanosecond-class pulsed visible laser is passively fanned out to the desired devices. Devices targeted by pump light are detuned far from resonance ($\Delta \gg \Gamma$) and effectively extinguished, whereas unactuated cavities retain their high $\Delta\approx 0$ reflectivity.

We used pump-probe spectroscopy for wide-field imaging of these few-nanosecond switching events (Appendix~\ref{appendix:setup}). Short infrared probe pulses were carved with a electro-optic amplitude modulator (DC biased to an intensity null) and variably delayed to coincide with the arrival of visible pump light at the PhC membrane, gating probe field transmission to the IR camera. We then measured the near- and far-field reflection as a function of the probe delay to reconstruct switching events with sub-nanosecond resolution. Fig.~\ref{fig:switching}e-f plots the resulting far-field intensity profiles $|E_r|^2$ for horizontal and vertical on-off gratings. For a $5$ ns probe pulse width, the maximum near-field extinction of targeted cavities (7.4 dB and 9.8 dB for horizontal and vertical gratings, respectively) occurs within a $\resim 6$ ns delay; i.e. just after the pump and probe pulses completely overlap. This minimum probe pulse width is limited by the requirement for high imaging contrast between probe pulses and leakage (due to the imperfect probe modulator extinction) given the instrument-limited trigger repetition rate ($\resim$MHz) and camera integration time.

As expected, the input field is primarily scattered into first-order diffraction peaks within the (greater than $10^\circ$) 2D field-of-view of $S(\vec{k})$. The illustrated cross sectional beam profiles again agree with analytic results for a 80\% filled linear array of uniform apertures (black dashed lines). For the horizontal grating Fig.~\ref{fig:switching}e, the fit is scaled by a factor of $\resim 2$ to account for the \textit{increased} reflectivity of unactuated cavities during switching events, which we attribute to residual coupling between adjacent cavities. In both cases, the pattern diffraction efficiencies --- measured as the fraction of integrated power within the outlined regions in Fig.~\ref{fig:switching} --- $(\eta_x, \eta_y) = (0.22, 0.20)$ compare favorably to the efficiency of the fitted uniform aperture array. Even with amplitude-dominant modulation, these metrics exceed the efficiencies of previous resonator-based experiments due to our high-directivity PhC antenna array \cite{Horie2017}. 

\section{Summary and Outlook}
These proof-of-concept experiments demonstrate near-complete spatiotemporal control of a narrow-band optical field filtered in space and time by an array of wavelength-scale, high-speed resonant modulators. While the general resonant architecture (Fig.~\ref{fig:spatiotemporal}c) is applicable to a range of microcavity geometries and modulation schemes, the combination of our high-$Q$, vertically-coupled PhC cavities with efficient, all-optical free-carrier modulation achieves (C1-5) with an ultrahigh per-pixel spatiotemporal bandwidth $\nu\approx 5.6~\text{MHz}\cdot\text{sr}$. This MHz-order modulation bandwidth per aperture-limited spatial mode corresponds to a more than ten-fold improvement over the 2D spatial light modulators reviewed in Fig~\ref{fig:spatiotemporal}b. Our wafer-scale fabrication and parallel trimming offer a direct route towards scaling this performance to spectrally-multiplexed, $\mathcal{O}(\text{cm}^2)$ apertures for exascale interconnects beyond the reach of current electronic systems, thus motivating the continued development of optical addressing and control techniques.

The PhC-SLM opens the door to a number of applications and opportunities, including: \textit{high-definition, high-frame-rate holographic displays} by the integration of a back-reflector (see Appendices~\ref{appendix:opt-overlap}-\ref{appendix:substrate}) for one-sided, phase-only, and full-DoF spatiotemporal modulation; \textit{compact device integration} via direct transfer printing of our cavity arrays onto a high-bandwidth $\upmu$LED array \cite{Carreira2020}; \textit{three-dimensional optical addressing and imaging} by combining on-demand $\upmu$LED control with statically trimmed detuning profiles that continuously steer pre-programmed patterns \cite{Shaltout2019}; \textit{large-scale programmable unitary transformations} for universal linear optics processors~\cite{Bogaerts2020};   \textit{focal plane array sensors} for high-spatial-resolution readout of refractive index perturbations in imaging applications from endoscopy to bolometry and quantum-limited superresolution \cite{Pahlevaninezhad2018,Watts2007,Grace2020}; \textit{optical neural network acceleration} via low-power, high-density unitary transformation of free-space optical inputs \cite{Hamerly2019,Wetzstein2020}; and \textit{high-speed adaptive optics} enabling free-space compressive sensing, deep-brain neural stimulation, and real-time scattering matrix inversion in complex media \cite{Mosk2012,Yoon2020}. Moreover, whereas we have so far considered only mode transformations, the PhC-SLM's high-$Q/V$ resonant enhancement suggests the possibility of programming the quantum optical excitations/fields of these modes for applications ranging from multimode squeezed light generation \cite{Bourassa2021}, to multiplexed single photon sources for linear optics quantum computing~\cite{Kok2007, Bartolucci2021} or deterministic photonic logic~\cite{Heuck2020,Krastanov2021}.

\begin{acknowledgments}

The authors thank Flexcompute, Inc. for supporting FDTD simulations, the MIT.nano staff for fabrication assistance, and M. ElKabbash (MIT) for useful discussions. C.P. was supported by the Hertz Foundation Elizabeth and Stephen Fantone Family Fellowship. S.T.M. is funded by the Schmidt Postdoctoral Award and the Israeli Vatat Scholarship. Experiments were supported in part by Army Research Office grant W911NF-20-1-0084, supervised by M. Gerhold, the Engineering and Physical Sciences Research Council (EP/M01326X/1, EP/T00097X/1), and the Royal Academy of Engineering (Research Chairs and Senior Research Fellowships). This material is based on research sponsored by Air Force Research Laboratory under agreement number FA8650-21-2-1000. The U.S. Government is authorized to reproduce and distribute reprints for Governmental purposes notwithstanding any copyright notation thereon. The views and conclusions contained herein are those of the authors and should not be interpreted as necessarily representing the official policies or endorsements, either expressed or implied, of the United States Air Force, the Air Force Research Laboratory or the U.S. Government.

C.P. and D.E. conceived the idea, developed the theory, and led the research. M.M. and C.P. developed the far-field optimization technique and designs. C.B. developed the optimized resonator detuning theory. C.P. conducted the experiments with assistance from I.C. (trimming experiments), S.T.M. (holography software), and A.G. ($\upmu$LED measurements). J.J.M., M.D., and M.S. contributed the $\upmu$LED arrays and guided the incoherent switching experiments. C.H. and J.W.B. fabricated the initial samples for evaluation prior to foundry process development by C.T., J.S.L., J.M., and G.L. S.P. assisted with wafer post-processing. M.F. coordinated and led the foundry fabrication with assistance from G.L. and D.C. C.P. wrote the manuscript with input from all authors. 

\end{acknowledgments}

\bibliography{ms-appendices}

\clearpage
\newpage
\appendix 

\setcounter{table}{0}
\renewcommand{\thetable}{A\arabic{table}}%
\setcounter{figure}{0}
\renewcommand{\thefigure}{A\arabic{figure}}%

\section{Analytic Model for Slab Switching}
\label{appendix:analytic}

Here, we develop an analytic model for all-optical switching in slab-type PhC cavities to estimate the required tuning energy. An absorbed control pulse produces a refractive index change
\begin{equation}
    \delta n(\vec{r},t) = -\alpha_c N(\vec{r},t) + \alpha_t T(\vec{r},t)
\end{equation}
proportional to the photo-excited carrier density $N$ and induced temperature change $T$ through the plasma dispersion and thermo-refractive effects, respectively. The thermo-refractive coefficient $\alpha_t = \d n/\d T$ and empirical free-carrier ``scattering volume" $\alpha_c = -\d n/ \d N$ are typically both positive such that the two effects counteract one another. The evolution of $\delta T$ and $N$ are governed by the diffusion equations
\begin{subequations}
\begin{align}
\frac{\partial N(\vec{r},t)}{\partial t} &= \nabla \cdot ( D_c\nabla N) - \frac{N}{\tau} + g(\vec{r},t) \\
\frac{\partial T(\vec{r},t)}{\partial t} &= \nabla \cdot (D_t\nabla T) + q(\vec{r},t)
\end{align}
\end{subequations}
given the thermal diffusivity $D_t$ and assuming ambipolar diffusion of carriers with lifetime $\tau$ and diffusivity $D_c$. Over relevant timescales $t > w^2/D_c$ in a $w$-thick uniform slab, vertical diffusion can be neglected to yield solutions
\begin{subequations}
\label{eqn:convolutions}
\begin{align}
    N(\vec{r},t) &= g(\vec{r},t) *_{\vec{r},t} G(\vec{r},2D_c t)e^{-t/\tau} \\
    T(\vec{r},t) &= q(\vec{r},t) *_{\vec{r},t} G(\vec{r},2D_t t)
\end{align}
\end{subequations}
expressed as convolutions ($*$) of the inhomogeneous sources $g(\vec{r},t)$ and $q(\vec{r},t)$ with the two-dimensional Green's function
\begin{equation}
    G(\vec{r},\sigma^2) = \frac{1}{2\pi w \sigma^2}\exp{-\frac{|\vec{r}|^2}{2\sigma^2}}.
\end{equation}
All variables are considered uniform along the vertical axis; $\vec{r}$ in our notation thus corresponds only to transverse coordinates in the slab plane. We specifically consider solutions to Eqns.~\ref{eqn:convolutions} in response to a focused, square-wave Gaussian control pulse with beam waist $2\sigma_p$, pulse-width $T$, and pulse (photon) energy $E$ ($E_0$) absorbed into the cavity with efficiency $\eta_\text{abs}$. The results can be considerably simplified with the conservative (i.e. underestimating plasma dispersion at short timescales $t 	\lesssim \tau$), albeit crude, assumption of instantaneous carrier diffusion to the diffusion length $\sqrt{D_c \tau}$. This method decouples carrier decay and diffusion to yield the carrier density
\begin{equation}
    N(\vec{r},t) = N_0(t)G(\vec{r},2D_c\tau + \sigma_p^2)
\end{equation}
with time-dependent total population
\begin{equation}
  N_0(t) = \eta_\text{abs}\frac{\tau}{T}\frac{E}{E_0}
  \begin{cases}
    (1-e^{-t/\tau}), & t \leq T \\
    e^{-t/\tau}(e^{T/\tau}-1), & t > T.
  \end{cases}
\end{equation}
The recombination of each carrier pair releases the bandgap energy $E_g$ back into the slab with volumetric heat capacity $c_v$, yielding the source
\begin{equation}
    q(\vec{r},t) = -\left(\frac{\partial N}{\partial t}\right)_\text{decay} \frac{E_g}{c_v} = \frac{N(\vec{r},t)}{\tau}\frac{E_g}{c_v}
\end{equation}
that produces the temperature profile
\begin{equation}
    T(\vec{r},t) = \frac{E_g}{ c_v \tau} N_0(t) *_t G(\vec{r},2D_tt+2D_c\tau + \sigma_p^2).
    \label{eqn:temp}
\end{equation}
Note that we neglect additional initial heating from above-band absorption.

Given sufficiently small $|\delta n|$, the resulting linewidth-normalized resonance shift
\begin{equation}
    \tilde{\Delta}(t) = \Delta(t)/\Gamma = - Q \int \frac{\delta n}{n}(\vec{r},t) |\vec{E}(\vec{r})|^2 \dV
    \label{eqn:pt}
\end{equation}
for the electric field profile $\vec{E}(\vec{r})$ with normalization ($\int |\vec{E}(\vec{r})|^2 \dV = 1$) is well-approximated by first-order perturbation theory \cite{Joannopoulos2008}. We consider a Gaussian-shaped mode envelope $|\vec{E}|^2 = G(\vec{r},\sigma_0)$ fully-confined with uniform transverse amplitude within the high-index slab. 

Since Eqn.~\ref{eqn:temp} must be evaluated numerically, we assume a constant temperature change $T(\vec{r},t)=T(0,t)$ across the mode --- valid for typical experimental regimes of interest where $\sigma_0 \ll 2D_tt + 2D_c\tau +\sigma_p$ --- to avoid the additional integration in Eqn.~\ref{eqn:pt}. The overlap between the optical mode and the static free carrier profile, on the other hand, can be analytically evaluated to yield the combined result
\begin{align}
    \tilde{\Delta}(t) = &\frac{\alpha_c}{n} N_0(t) \frac{\sigma_0^2}{2D_c\tau + \sigma_p^2+ \sigma_0^2} \left( \frac{Q}{V} \right)  \nonumber \\ 
    &-Q\frac{\alpha_t}{n}\frac{E_g}{ c_v \tau} N_0(t) *_t G(0,2D_tt+2D_c\tau+\sigma_p^2)
\end{align}
for the cavity mode volume $V=\int \epsilon |E|^2 \dV /\max{\lbrace \epsilon |E|^2\rbrace }=2\pi w \sigma_0^2$. Since the reflected signal directly tracks the cavity amplitude in cross-polarization, the normalized reflectivity
\begin{equation}
    r(t) =  \int_0^t \d t' e^{-\Gamma(t-t')-i\int_{t}^{t'} \d t'' 2\Delta(t'')}
\end{equation}
is finally found by numerically integrating the cavity evolution as dictated by coupled mode theory \cite{Haus1984}. 

\begin{figure}
\includegraphics[width=0.42\textwidth]{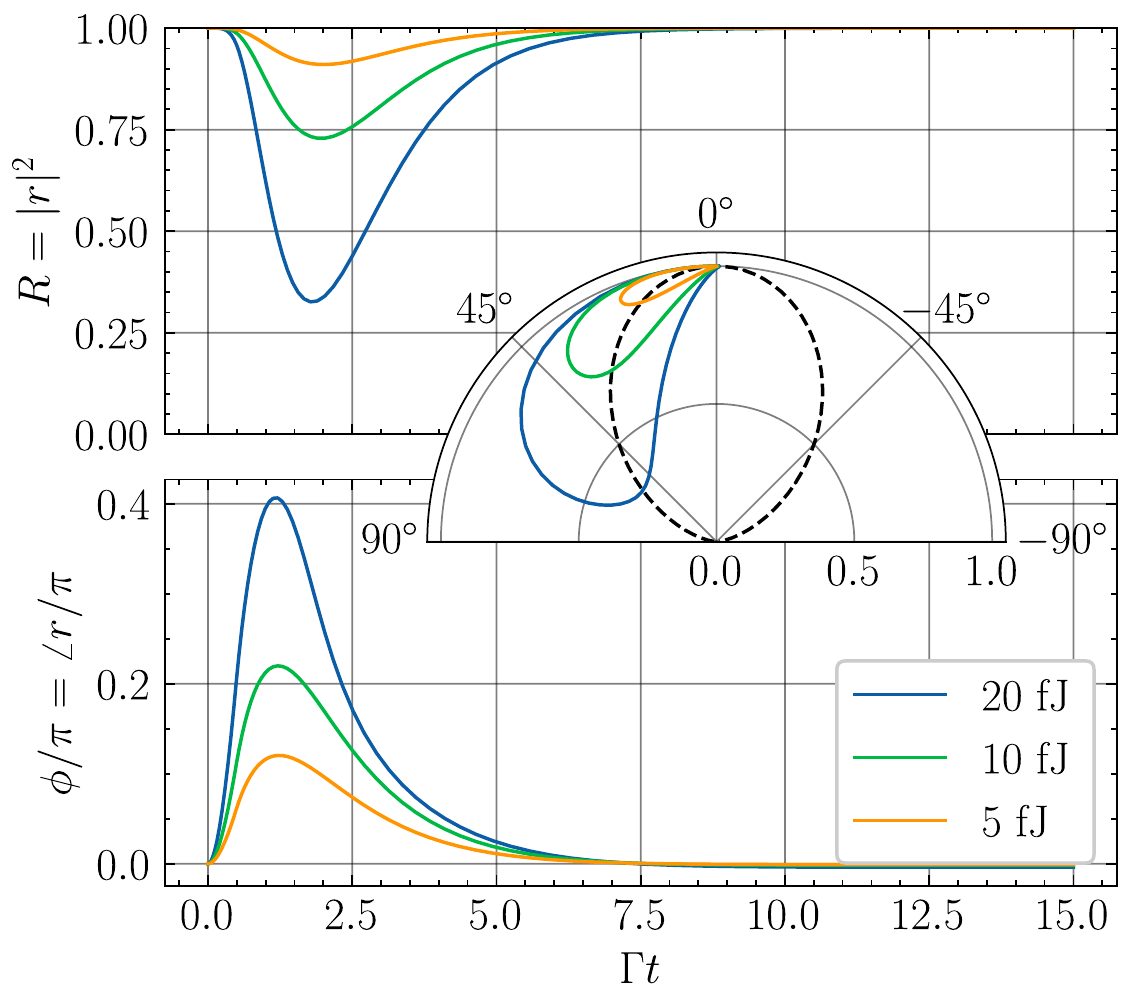}
\caption{Estimated normalized reflection coefficient $r(t)=\sqrt{R(t)}e^{j\phi(t)}$ as a function of switching energy for the parameters in Table~\ref{table:parameters}. Insets: same results in polar form for comparison to the Lorentzian reflection profile $r(\Delta)=1/(1+j\Delta)$ of a cavity under cross-polarized excitation with static detuning $\Delta$ (black dashed line).}
\centering
\label{figS:switching}
\end{figure}

Fig.~\ref{figS:switching} plots the switching characteristics for the parameters in Table~\ref{table:parameters}. Free-carrier dispersion dominates the response for the nanosecond-order timescales of interest followed by a slow ($\upmu$s-order), weak ($|\tilde{\Delta}| \ll 1$) thermal rebound. The three order-of-magnitude timescale difference effectively decouples the two modulation mechanisms. Note that the true reflection coefficient deviates from the Lorentzian response of a quasi-static cavity due to the fast (relative to the cavity decay rate $\Gamma$) carrier dynamics. These results indicate that a SLM with $10^6$ pixels operating with $\omega_s>2\pi\times 100$ MHz could be realized with $\mathcal{O}(\text{watt})$ optical control power.

\renewcommand{\arraystretch}{1.2}
\begin{table}
\footnotesize
\centering
\vspace{10pt}
\begin{tabular}{@{} C{1in}C{1in}C{1in} @{}}
\toprule
\textbf{Parameter} & \textbf{Value} & \textbf{Source} \\
\midrule
$n_\text{Si}$ & 3.48 & \cite{Komma2012} \\
$E_g$ & 1.12 eV & \cite{Komma2012} \\
$\alpha_t$ & $1.8\times 10^{-4}$ K$^{-1}$ & \cite{Komma2012} \\ 
$c_{v}$ & 1.64 J/cm$^3\cdot$K & \cite{Panuski2020} \\
$D_t$ & 0.26 cm$^2$/s & \cite{Panuski2020} \\
$\alpha_c$ & $8\times 10^{-9}$ $\upmu$m$^{3}$ & \cite{Nedeljkovic2011} (Linearized)\\ 
$D_c$ & 19 cm$^2$/s & \cite{Tanabe2008} \\
$\tau$ & 1 ns & \cite{Tanabe2008} \\

$\lambda_0$ & 1550 nm & Assumed \\ 
$Q$ & 200,000 & Assumed \\ 
$\tilde{V}$ & 0.95 & \cite{Panuski2020} \\ 
$w$ & 220 nm & Measured \\
$2\sigma_0$ & $0.66~\upmu\text{m}$ & $2\sqrt{V/(2\pi w)}$ \\

$\lambda_p$ & $0.53~\upmu\text{m}$ & Assumed \\
$T$ & 0.5 ns & Assumed \\
$2\sigma_p$ & $0.22~\upmu\text{m}$ & $\lambda_p/2$ \\
$\eta_\text{abs}$ & 0.6 & FDTD \\

\bottomrule
\end{tabular}
\caption{Parameters used for the simulated switching results in Fig.~\ref{figS:switching}. The pulse parameters were selected to mimic the typical experimental conditions of Section~\ref{sec:switching} in the main text.}
\label{table:parameters}
\end{table}

\section{Performance Comparisons}
\label{appendix:litreview}

Table~\ref{table:slms} compares the PhC-SLM demonstrated here to other actively-controlled, 2D SLMs (Fig.~\ref{fig:spatiotemporal}b). Wavelength-steered devices and switch arrays are omitted to restrict focus to the typical SLM architectures in  Fig.~\ref{fig:spatiotemporal}. Notably, while beamsteering with PhC waveguides \cite{Ito2020,Vercruysse2021,Tamanuki2021} and laser arrays \cite{Sakata2020} has recently been demonstrated, our device is the first (to our knowledge) to feature simultaneous emission from a 2D array of individually controllable PhC pixels.

\renewcommand{\arraystretch}{1.2}
\begin{table*}
\footnotesize
\centering
\vspace{10pt}
\begin{tabular}{@{} C{2cm}C{5.5cm}C{2cm}C{3cm}C{1cm}C{2cm} @{}} \toprule
\bf Class [Year] & \bf Device & $\bm{N_x\times N_y}$ & $\bm{\Omega_s = \frac{\lambda}{\Lambda_x} \times \frac{\lambda}{\Lambda_y}}$ & $\bm{\zeta}$ \bf{[$\bm{\%}$]} & $\bm{\omega_s/2\pi}$ \bf{[Hz]} \\
\midrule

\bf EO [2022] & \bf PhC-SLM & $\bm{8\times 8}$ & $\bm{10.6^\circ \times 14.5^\circ}$ & \bf 64 & $\bm{1.4\times 10^8}$ \\

EO [2021] & $\chi^{(2)}$ polymer-coated grating \cite{Benea2021} & $2\times 2$ & $0.2^\circ \times 0.2^\circ$ & --- & $5.0 \times 10^7$ \\

EO [2019] & $\chi^{(3)}$ thin-film plasmonic resonator \cite{Smolyaninov2019} & $4\times 4$ & $0.8^\circ \times 1.1^\circ$ & 20* & $1.0\times 10^9$ \\

EO [2017] & Bilayer guided resonators \cite{Shuai2017} & $6\times 6$ & $0.3^\circ \times 0.3^\circ$ & 40* & $2\times 10^8$ \\

EO [2011] & $\chi^{(2)}$ polymer-coated grating \cite{Greenlee2011} & $4\times 4$ & $0.1^\circ \times 0.1^\circ$ & 18* & $8.0 \times 10^5$ \\

EO [2005] & MQW micropillar modulators \cite{Junique2005} & $128\times 128$ & $1.3^\circ \times 1.3^\circ$ & 50 & $1.3\times 10^7$ \\

\midrule


Thermal [2018] & Asymmetric Fabry-Perot cavity \cite{Horie2017} & $6\times 6$ & $3.4^\circ \times 3.4^\circ$ & 59 & $1.4\times 10^4$ \\

Thermal [2013] & Waveguided phased array \cite{Sun2013,Yaacobi2014} & $8\times 8$ & $9.9^\circ \times 9.9^\circ$ & 10* & $1.1\times 10^5$ \\

\midrule


MEMS [2019] & Grating phase shifters \cite{Wang2019} & $160\times 160$ & $4.4^\circ \times 4.1^\circ$ & 85* & $5.5\times 10^4$ \\

MEMS [2019] & Piston mirrors \cite{Bartlett2019} & $960\times 540$ & $3.4^\circ \times 3.4^\circ$ & --- & $2.0\times 10^4$ \\ 

MEMS [2014] & High-contrast gratings \cite{Yang2014} & $8\times 8$ & $2.7^\circ \times 2.7^\circ$ & 36* & $5.0\times 10^5$ \\

MEMS [2001] & Piston mirrors \cite{Shrauger2001} & $256\times 256$ & $0.9^\circ \times 0.9^\circ$ & 86 & $5.0\times 10^5$ \\

\midrule

LC [2020] & Plasmonic metasurface \cite{Li2020} & $3\times 2$ & $0.3^\circ \times 0.3^\circ$ & --- & $2.5\times 10^1$ \\

LC [2019] & ``MacroSLM" \cite{Marshel2019} & $1536\times 1536$ & $3.0^\circ \times 3.0^\circ$ & 95 & $6.0\times 10^2$ \\

LC [1994] & Binary ferroelectric LC \cite{McKnight1994} & $256\times 256$ & $2.2^\circ \times 2.2^\circ$ & 79 & $8.3\times 10^3$ \\
\bottomrule
\end{tabular}
\caption{Performance comparison of selected active 2D spatial light modulators from  Fig.~\ref{fig:spatiotemporal}b. Estimated fill factors $\zeta$ are marked by a *.}
\label{table:slms}
\end{table*}

\section{Inverse Design Strategy}
\label{appendix:gme}

We implement the inverse design strategy in Fig.~\ref{fig:holograms}b using the open-source guided mode expansion (GME) package \texttt{Legume} \cite{Minkov2020}. GME approximates the cavity eigenmode using the incomplete basis set of waveguide modes in an ``effective" unpatterned slab (in effect transforming the 3D eigenproblem to 2D) and perturbatively computes the loss due to coupling to the radiative continuum \cite{Andreani2006}. 

During each optimization step, we aggregate the losses of the fundamental slab mode over four Bloch boundary conditions $\vec{k}_i$ at all wave vectors  $\vec{g}_{mn}=\vec{k}_i+\vec{G}_{mn}=\vec{k}_i+2\pi(\frac{m}{\Lambda_x},\frac{n}{\Lambda_y})$ satisfying $|\vec{g}_{mn}|<g_\text{max}=2.5\times 2\pi/a$ given the reciprocal lattice vectors $\vec{G}_{mn}$. The objective function Eqn.~\ref{eqn:objective} converges within tens of iterations, and the resulting design is then verified with $g_\text{max}=3\times 2\pi/a$ using a $3\times3$ $\vec{k}_i$ grid in the Brillouin zone of the rectangular lattice of unit cells. 

\begin{figure}
\includegraphics[width=0.42\textwidth]{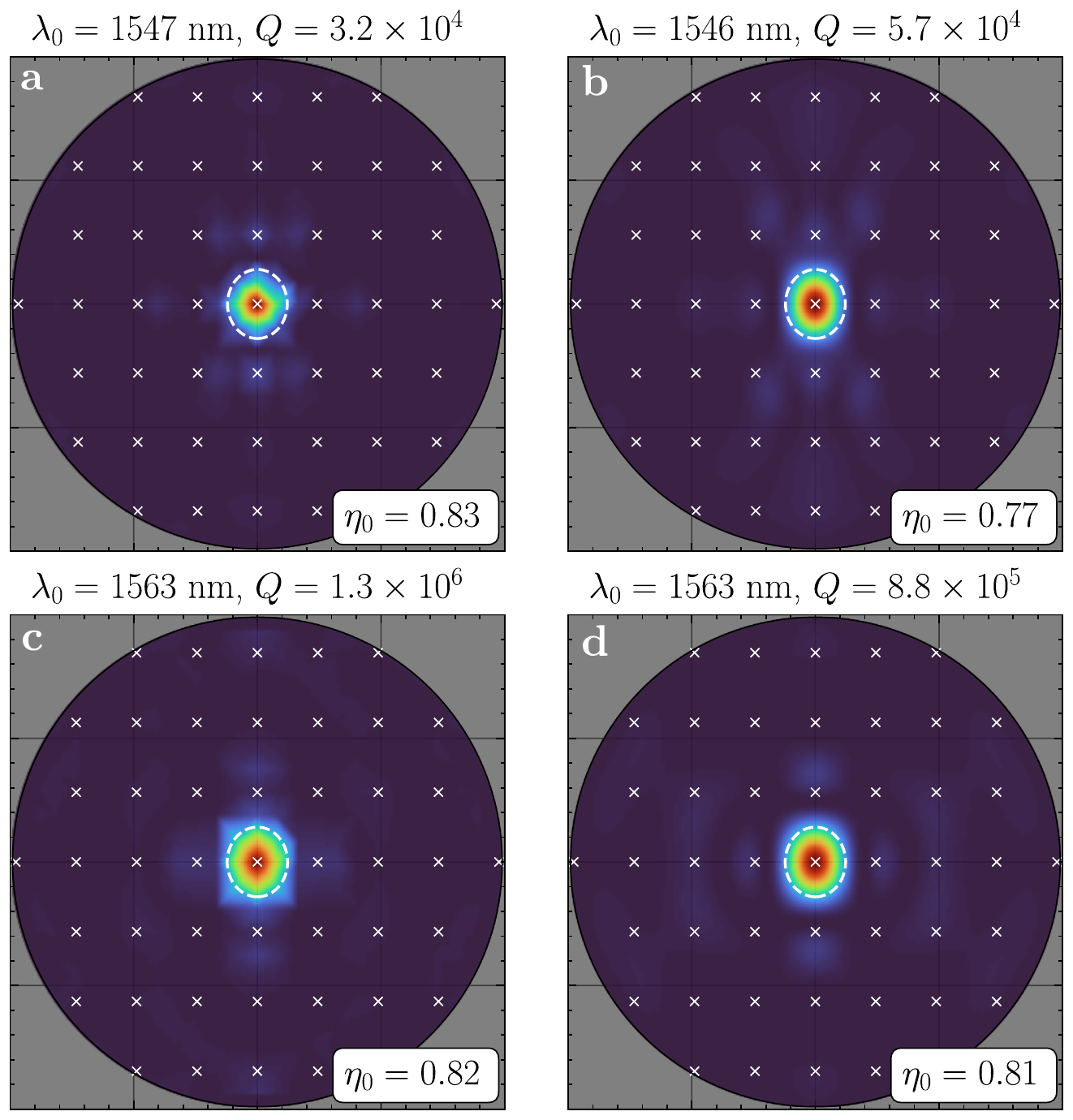}
\caption{Comparison of resonant wavelengths $\lambda_0$, quality factors $Q$, and far-field emission spectra computed from GME (left) and FDTD (right) simulations for two target $Q_0$ (top, bottom).}
\centering
\label{figS:gmeVfdtd}
\end{figure}

Exemplary GME-approximated far-field profiles for an $L3$ cavity with two target quality factors $Q_0$ are shown in Fig.~\ref{figS:gmeVfdtd}a,c for comparison to those computed using near-to-far-field transformations of FDTD-simulated fields \cite{Vuckovic2002,Kim2006}. These results confirm that the perturbatively-computed GME coupling coefficients can be used to accurately estimate a cavity's far-field scattering profile.  

\begin{figure}
\includegraphics[width=0.48\textwidth]{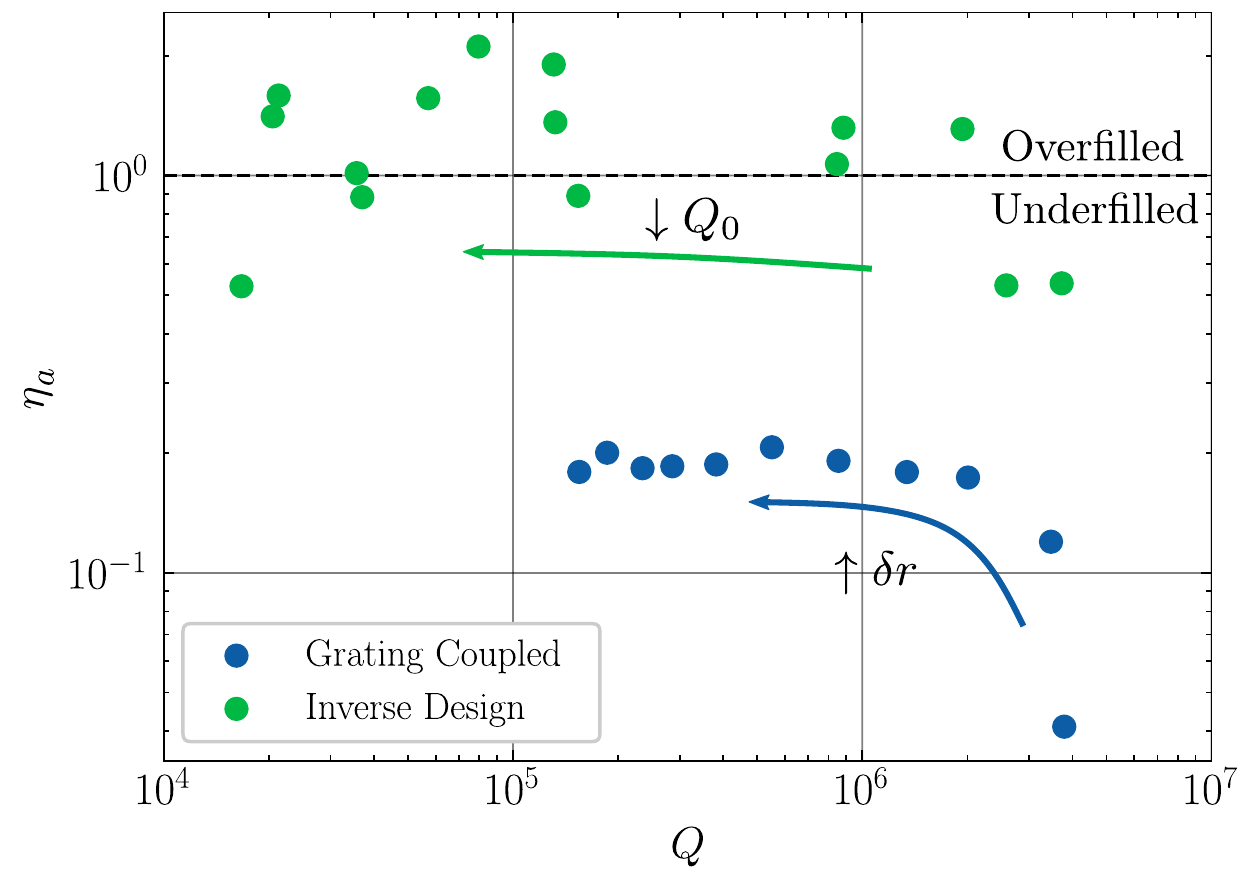}
\caption{FDTD-computed aperture efficiency $\eta_a$ of vertically coupled $L3$ cavities based on a grating perturbation $\delta r/r \in [0, 0.05]$ or an inverse-design target quality factor $Q_0\in [10^2, 10^6]$. Inset arrows illustrate parameter trends.}
\centering
\label{figS:apertureEff}
\end{figure}

The inverse design objective function (Eqn.~\ref{eqn:objective}) maximizes the directivity $D=4\pi S(0)/\int_\Omega S(\vec{k})~\d\Omega$ (for the light cone $\Omega$) of the emission profile $S(\vec{k})$ for any $Q_0$. The resulting aperture efficiency 
\begin{equation}
    \eta_a = \frac{D_0}{\max D} = \frac{\lambda^2}{A}\frac{S(0)}{\int_{\Omega} S(\vec{k})~\d\Omega } = \frac{A_0}{A}
\end{equation}
compares $D$ to the maximum directivity $4\pi A/\lambda^2$ of an area $A=\Lambda_x\Lambda_y$ aperture at wavelength $\lambda$, and can therefore be interpreted as the fill factor of light scattered from an effective area $A_0$. Fig.~\ref{figS:apertureEff} compares $\eta_a$ for grating-coupled and inverse-designed $L3$ cavities. For most inverse designs, $\eta_a\approx 1$ regardless of $Q_0$. Since GME assumes periodic boundary conditions (indicative of the true array design), scattering from neighboring unit cells enables designs with $\eta_a>1$. However, this ``super-directive" performance is undesirable since the steerable field-of-view is narrowed to $\Omega_p < \Omega_s$.

\section{Experimental Setups}
\label{appendix:setup}

\begin{figure}
\includegraphics[width=0.42\textwidth]{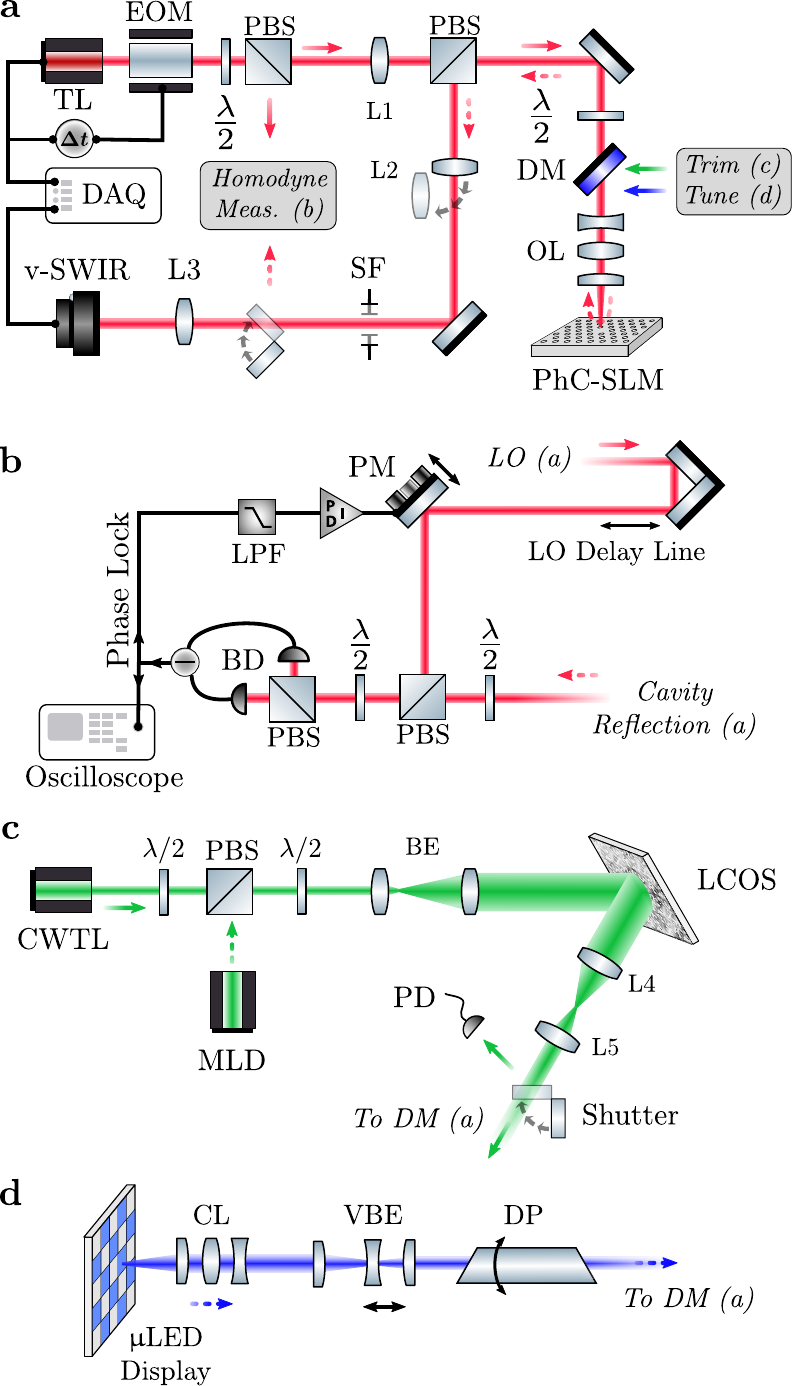}
\caption{Overview of experimental setups for measuring and controlling the photonic crystal SLM (PhC-SLM). A cross-polarized microscope (a) featuring balanced homodyne measurement (b) enables near- and far-field characterization of cavity arrays controlled by SLM-distributed coherent light (c) or high-speed incoherent $\upmu$LED arrays (d). TL: tunable infrared laser (Santec TSL-710), EOM: electro-optic amplitude modulator;  $\lambda/2$: half-wave plate, PBS: polarizing beamsplitter; L1: 250 mm back-focal-plane lens; DM: long-pass dichroic mirror; OL: objective lens (Nikon Plan Fluor 40$\times$/0.60 NA or Nikon LU Plan 100$\times$/0.95 NA), L2: 250 mm back-focal-plane lens; SF: spatial filter; L3: 200 mm tube lens; v-SWIR: visible-short wave infrared camera (Xenics Cheetah 640); DAQ: data acquisition unit (NI USB-6343); $\Delta t$: trigger delay generator (SRS DG645); LO: local oscillator; PM: piezo mirror; BD: balanced detector (Thorlabs PDB480C-AC); Phase Lock: TEM LaseLock; LPF: low-pass filter; CWTL: continuous-wave trimming laser (Coherent Verdi V18); MLD: modulated laser diode (Hubner Cobolt or PicoLAS LDP); BE: $5\times$ visible beam expander; LCOS: high-power liquid crystal SLM (Santec SLM-300); L4: 300 mm; L5: 250 mm; PD: photo-detector; CL: collection lens (Zeiss Fluar 5$\times$/0.25 NA); VBE: $0.5\times-2\times$ variable beam expander; DP: dove prism.}
\centering
\label{figS:setup}
\end{figure}

Fig.~\ref{figS:setup} schematically illustrates the major components of our experimental setup. Here, we describe the design and function of each sub-assembly.

\subsection{Near-Field Reflection Spectra}
The wide-field, cross-polarized microscope in Fig.~\ref{figS:setup}(a) allows us to simultaneously measure the reflection from every cavity within a camera's field-of-view. A visible illumination path (not illustrated) is joined with collimated infrared light from a tunable laser with a dichroic mirror and focused onto the back-focal-plane (BFP) of an objective by lens L1. The angle-of-incidence and spot size of the infrared beam on the sample are therefore controlled by translating L1 and varying the collimated beam diameter, respectively. In our typical wide-field configuration, a 7.2 mm beam diameter focused to the center of a $40\times$ objective's BFP yields a $\resim 150 ~\upmu$m waist-diameter, vertically-incident field that quasi-uniformly illuminates $10\times 10$ PhC cavity arrays.

By orienting the input polarization at a $45^\circ$ angle relative to the dominant cavity polarization axis (with a half-wave plate or by physically rotating the sample), light coupled into and reflected by the PhC cavity is polarization rotated and can be isolated from direct, specular reflections with a polarizing beamsplitter. A kHz-rate free-running, dual-band (visible and infrared) camera images this cross-polarized reflection signal through the tube lens L3. For each frame collected during a laser sweep, the wavelength is interpolated from the recorded camera and laser output triggers and each cavity's reflection is integrated over a fraction of pixels within its imaged unit cell boundary. We use the resulting high-contrast reflection spectra (across all devices within the field-of-view) to characterize device performance and monitor the cavity trimming process. 

The sample mount below the objective (OL) is temperature stabilized to within $10$ mK with a Peltier plate and feedback controller. For trimming experiments, the sample is placed in a high-pressure oxygen environment within a custom chamber offering in-situ optical access through a glass window.   

\subsection{Calibrated Far-Field Measurement}
\begin{figure}
\includegraphics[width=0.42\textwidth]{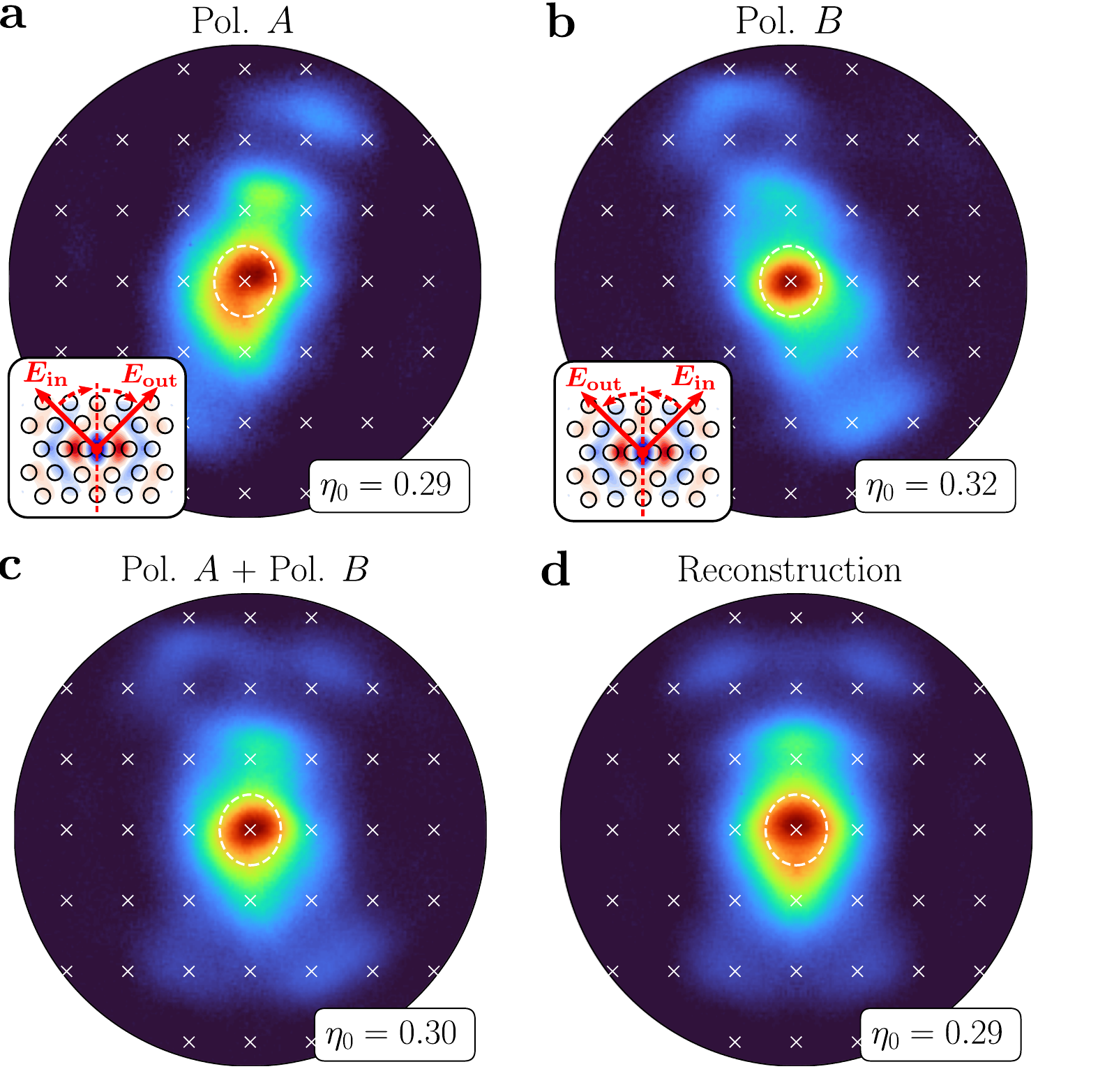}
\caption{Cross-polarized back-focal-plane (BFP) imaging techniques for a grating-coupled $L3$ cavity. Two orthogonally polarized far-field profiles are imaged by orienting the input polarization $E_\text{in}$ at a $+45^\circ$ (a) or $-45^\circ$ (b) angle from the dominant cavity polarization axis (dashed line in inset). The complete cavity emission profile $S(\vec{k})$ can be reconstructed by summing both images (c) or approximated from a single polarized image (d), yielding near-identical images with quantitative agreement between the extracted $\eta_0$.}
\centering
\label{figS:bfpmethod}
\end{figure}

Inserting a lens (L2) in the collection path one focal length from the objective BFP allows us to measure the far-field profile $S(\vec{k})$ of individual or multiple cavities using the same setup. We position an iris at the intermediate image plane --- located with a removable lens (not shown) placed before L3 --- to spatially filter the emission from desired devices. We also calibrate the BFP scale using a reflective reference grating with known pitch. Due to the cross-polarized configuration, only a single polarization $\tilde{S}(\vec{k})\big|_{\theta}$ is imaged for any cavity-input polarization angle difference $\theta$. The complete cavity emission profile 
\begin{equation}
    S(\vec{k}) = \tilde{S}(\vec{k})\big|_{\theta}+\tilde{S}(\vec{k})\big|_{\theta\pm\pi/2}
\end{equation}
can therefore be reconstructed by sequentially imaging both polarizations as in Figs.~\ref{figS:bfpmethod}a-c for $\theta=45^\circ$. For maximum accuracy, we used this technique for the experimental results in Fig.~\ref{fig:singlecav}.

Alternatively, the specific choice $\theta=45^\circ$ allows $S(\vec{k})$ to be reconstructed from a single measurement. Due to mirror symmetry about the cavity's principal polarization axis $\hat{y}$, Fig.~\ref{figS:bfpmethod}a-b show that $\hat{\sigma}_{\hat{y}}\lbrace \tilde{S}(\vec{k})\big|_{\pm 45^\circ}\rbrace=\tilde{S}(\vec{k})\big|_{\mp 45^\circ}$ for the reflection operator $\hat{\sigma}$. This alternative reconstruction
\begin{equation}
    S(\vec{k}) = \left[ 1+\hat{\sigma}_{\hat{y}} \right] S(\vec{k})\big|_{\pm 45^\circ}
\end{equation}
is demonstrated experimentally in Fig.~\ref{fig:singlecav}d, yielding excellent agreement with Fig.~\ref{fig:singlecav}c. This technique simplifies high-throughput far-field measurements across cavity arrays (Fig.~\ref{figS:bfp-uniformity}, for example). 

\subsection{Homodyne Measurement}
The shot-noise-limited balanced homodyne detection setup in Fig.~\ref{figS:setup}c enables complex reflection coefficient measurements with greater than $>$3 dB shot-noise clearance below $1$ GHz \cite{Panuski2020}. Signal light reflected from the cavity combines with a path-length-matched (to within $\resim$mm based on time-delay measurements with a picosecond-class pulsed laser) local oscillator (LO), and both signals are coupled into a balanced detector using anti-reflection coated fibers. The in-phase ($I(t)$) and quadrature ($Q(t)$) components of the cavity reflection were sequentially measured by locking to the first and second harmonics of the balanced output in the presence of a piezo-driven LO phase dither. The resonant, cross-polarized cavity reflection $R$ and phase shift $\phi$ are then reconstructed as
\begin{align}
    R = \frac{\left[V_p - I(t)\right]^2 + Q^2(t)}{V_p^2} && \phi = \arctan\frac{Q(t)}{V_p - I(t)}
\end{align}
by normalizing to the measured peak voltage swing $V_p$ of the interference signal.

\subsection{Parallel Cavity Trimming}
A liquid crystal on silicon (LCOS) SLM (Fig.~\ref{figS:setup}c) actively distributes a high-power, continuous-wave visible laser to target devices during the cavity trimming procedure. The input laser was tunably attenuated with a motorized half-wave plate (preceding a PBS) and subsequently expanded to overfill the LCOS aperture. The LCOS SLM was re-imaged onto the objective BFP (as confirmed by imaging with L2 in place) using two lenses (L4, L5) with focal lengths chosen to optimally match the imaged SLM and objective pupil dimensions. Phase retrieval-computed holograms then evenly distribute power to an array of focused spots on the sample (Appendix~\ref{appendix:qpslm}) when the mechanical, flip mirror shutter is opened.

\subsection{\texorpdfstring{$\upmu$}{u}LED Imaging}
The collection optics in Fig.~\ref{figS:setup}d maximize the intensity of a $\upmu$LED array projected onto the PhC membrane within the constraints dictated by the constant radiance theorem of incoherent imaging. Assuming a Lambertian emission profile, geometric optics gives the collection efficiency $\eta_c = \alpha_c^2$ for an objective lens (CL) with numerical aperture $\alpha_c$ focused on the $\upmu$LED array. The projection efficiency $\eta_p$ through the projection objective (OL, with numerical aperture $\alpha_p$) depends on the relative pupil sizes of both objectives and can be similarly approximated from geometric optics. The resulting intensity enhancement $\zeta=\eta_c\eta_p/M^2$ between the source and image (with magnification $M$) reaches a maximum $\zeta_\text{max} = \frac{1}{M^2+(1-\alpha_p^2)/\alpha_p^2}$ when the CL-collimated light overfills the back aperture of OL. The resulting design criteria,
\begin{equation}
     \alpha_c > \sqrt{\frac{M^2\alpha_p^2}{(M^2-1)\alpha_p^2+1}},
     \label{eqn:collectionNA}
\end{equation}
is achieved for our imaging setup with $\alpha_c = 0.25$, $\alpha_p = 0.95$, and $M\approx 1/30$. After CL, The overall magnification and rotation are fine-tuned with a variable beam expander and Dove prism, respectively.

\section{Optimum Gaussian Coupling}
\label{appendix:opt-overlap}

The far-field spatial overlap integral \cite{Munsch2013}
\begin{equation}
    O(w_0) = R^2 \iint_\Omega \vec{E}_c(\theta,\phi) \times \vec{H}_g^*(w_0,\theta,\phi) \sin \theta d\theta d\phi 
\end{equation}
over the hemisphere $\Omega$ at distance $R$ yields the power coupling $|O(w_0)|^2$ between a cavity mode $c$ and fundamental Gaussian beam $g$ as a function of the Gaussian waist radius $w_0$. We compute the cavity electric field profile $\vec{E}_c(\theta,\phi)$ by applying a near-to-far field transformation to the FDTD-simulated cavity mode in a plane just above and parallel to the PhC slab \footnote{Flexcompute, Inc. Tidy3D. \url{https://simulation.cloud}} \cite{Hughes2021}. The resulting far-field profile is finely discretized --- relative to the diffraction-limited beamwidth $\lambda/\Lambda \approx 14^\circ$ for typical cavity unit cell dimensions $\Lambda$ and resonant wavelength $\lambda$ ---  on a 1$^\circ$ grid in zenith and azimuth ($\theta$ and $\phi$, respectively). For the co-polarized Gaussian mode, we instead convert the magnetic field \cite{Haus1984}
\begin{align}
    \vec{H}(x,y,z) = &\sqrt{\frac{\pi}{Z_0}}\frac{2w_0}{z+j\pi w_0^2}\exp{\frac{-j2\pi(x^2+y^2)}{2(z+j\pi w_0^2)}} \nonumber \\
    \times &\left[\hat{x} + \left(\frac{-xz-jx\pi w_0^2}{z^2+(\pi w_0^2)^2}\right)\hat{z}\right] e^{-j2\pi z}
\end{align}
derived from paraxial diffraction theory (for the free space impedance $Z_0$) to spherical coordinates at a far-field distance $R$. Both modes are normalized to carry unit power $\int_\Omega \Re \lbrace \vec{E}\times \vec{H}^*\rbrace/2$.

\begin{figure}
\includegraphics[width=0.46\textwidth]{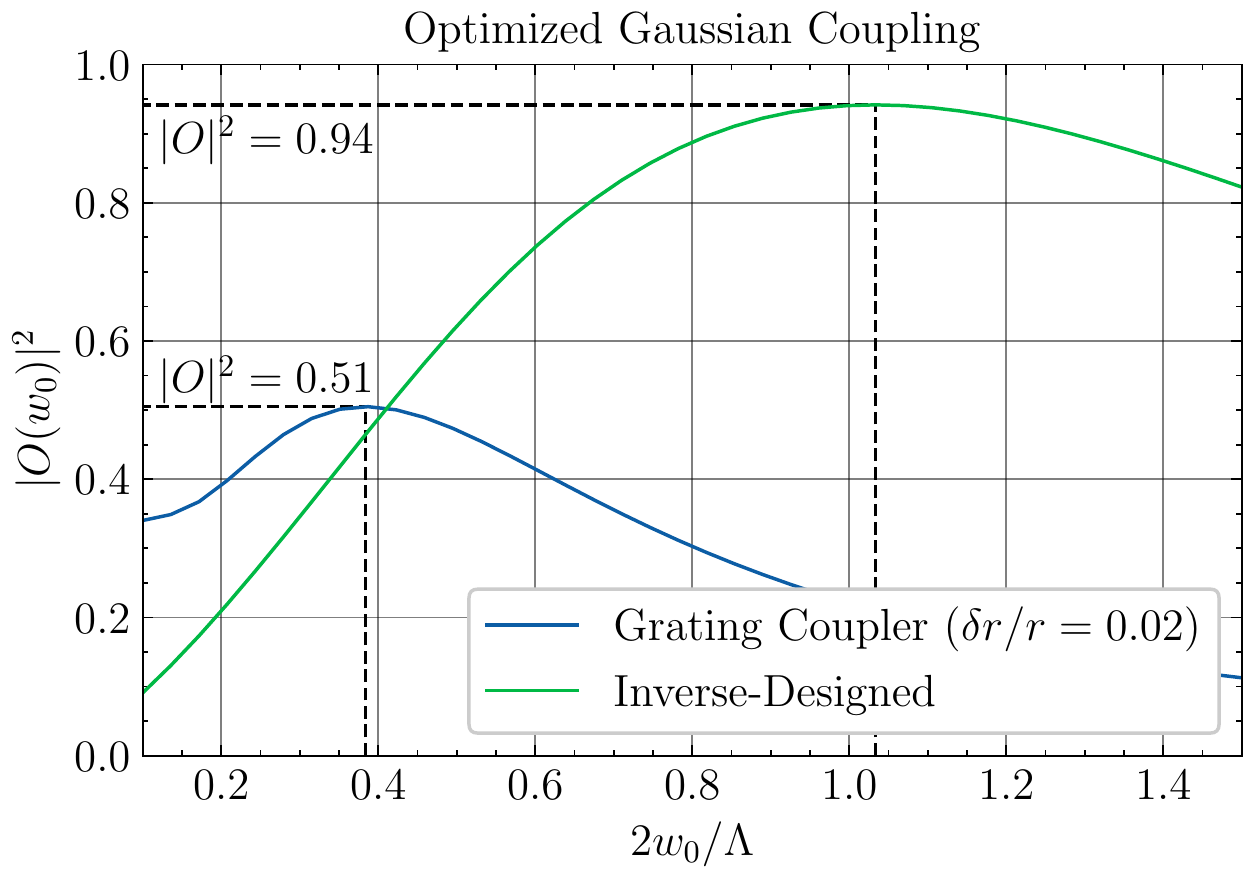}
\caption{Comparison of coupling between a fundamental Gaussian beam with waist $w_0$ and two different cavity designs with maximum unit cell dimension $\Lambda$.}
\centering
\label{figS:optcoupling}
\end{figure}

Fig.~\ref{figS:optcoupling} compares the resulting power coupling as a function of normalized waist $2w_0/\Lambda$ for grating-coupled and inverse-designed cavity designs with maximum dimension $\Lambda$. In both cases, a backreflector is assumed for unidirectional emission into $\Omega$. Besides the increase in maximum coupling to 94\%, the optimized waist diameter $2w_0\approx \Lambda$ indicates that the inverse-designed cavity's effective near-field scattering profile fully fills the design unit cell.  

The power coupling $|O|_\text{max}^2$ to a single desired free-space mode also allows us to compute that mode's amplitude reflection spectrum \cite{Haus1984}
\begin{equation}
    r(\tilde{\Delta}) = \frac{2|O|_\text{max}^2/p-j\tilde{\Delta}}{1+j\tilde{\Delta}}
    \label{eqn:r(delta)}
\end{equation}
using temporal coupled mode theory assuming a normalized detuning $\tilde{\Delta}=\Delta/\Gamma$ from the cavity resonance and $p$-directional emission (i.e. $p=1$ for unidirectional emission with a backreflector or $p=2$ for symmetric emission). Fig.~\ref{fig:singlecav} of the main text plots these optimized spectra. Whereas the grating-coupled cavity is undercoupled with an amplitude-dominant reflection spectrum, the inverse-designed cavity is phase-dominant as desired for high efficiency beamforming.

\section{Beamforming with a Coupled Amplitude-Phase Response}
\label{appendix:convex}
Phase retrieval algorithms (Appendix~\ref{appendix:qpslm}) optimize unity-magnitude near-field reflection coefficients $r=e^{j\phi}$ to generate a desired far-field intensity pattern, but fail for the coupled amplitude and phase reflection coefficients of Eqn.~\ref{eqn:r(delta)} \cite{Levi1984, Lesina2020}. Here, we derive an alternative algorithm to compensate for this coupling.

Assuming a spatially uniform field $E_i(\vec{r})=E_i$ incident on a $\Lambda_x \times \Lambda_y$-pitch array, the reflected far-field defined in Eqn.~\ref{eqn:far-field} is the product of the single-element far-field pattern $S(\vec{k})$ and the array factor
\begin{equation}
    \mathcal{F}\lbrace r \rbrace = \sum_{m,n} \Gamma_{m,n} \exp{j k (m \Lambda_x u + n \Lambda_y v)},
    \label{eqn:AF}
\end{equation}
i.e., the 2D discrete Fourier transform ($\mathcal{F}$) of the reflection coefficients $r$ with respect to the cosine-space coordinates $u=k_x/k$ and $v=k_y/k$. Since $S(\vec{k})$ is determined by element design (Section~\ref{sec:qff}), far-field pattern synthesis is achieved by optimizing the array factor. Given Eqn.~\ref{eqn:AF}, we can represent constrained far-field synthesis with the nonlinear program
\begin{equation}
\min_{\Delta} f(\Delta)= \min_{\Delta} \frac{1}{2} \norm{\frac{|\mathcal{F}\{r(\Delta)\}|^2}{\norm{|\mathcal{F}\{r(\Delta)\}|^2}_F} - \frac{|\textbf{Y}|^2}{\norm{|\textbf{Y}|^2}_F}}_F^2,
\label{eqn:NLProg}
\end{equation}
where $r(\Delta)$ is the matrix of reflection coefficients as a function of the detuning matrix $\Delta$, $|\cdot|$ represents the complex modulus, $\norm{\cdot}_F$ is the Frobenius norm, and $Y$ is the goal image. By normalizing the Fourier transforms, this program optimizes image appearance as opposed to absolute intensity, which varies in the presence of reflection loss. 

We approximate the solution to the nonlinear program using the L-BFGS-B method --- a low-memory quasi-Newton method with simple box constraints to limit detunings \cite{Liu1989}. L-BFGS-B uses the gradient and a low-rank approximation of the objective function's Hessian to approximate Newton's method, yielding superlinear convergence.

To efficiently optimize Eqn.~\ref{eqn:NLProg}, we computed its analytic gradient
\begin{align}
\nabla f = &\frac{2\left( g(a_1,b_1) - \frac{ \langle |\mathcal{F}\{r(\Delta)\}|^2, h\rangle_F }{\norm{|\mathcal{F}\{r(\Delta)\}|^2}_F}  g(a_2,b_2) \right)}{\norm{|\mathcal{F}\{r(\Delta)\}|^2}_F^2},
\label{eqn:gradient}
\end{align}
for
\begin{align}
    g(a, b) = \big[&(\Re{\mathcal{F}\{a\}} + \Im{\mathcal{F}\{b\}})\odot \Re{\nabla r}  + \nonumber\\
    &(\Re{\mathcal{F}\{b\}} -  \Im{\mathcal{F}\{a\}})\odot \Im{\nabla r} \big], \nonumber
\end{align}
\begin{equation}
    h = \frac{|\mathcal{F}\{r(\Delta)\}|^2}{\norm{|\mathcal{F}\{r(\Delta)\}|^2}_F} - \frac{|\textbf{Y}|^2}{\norm{|\textbf{Y}|^2}_F}, \nonumber
\end{equation}
and 
\begin{align}
a_1 = h \odot \Re{\mathcal{F}\{r(\Delta)\}} && a_2 = |\mathcal{F}\{r(\Delta)\}|^2 \odot \Re{\mathcal{F}\{r(\Delta)\}} \nonumber \\
b_1 = h \odot \Im{\mathcal{F}\{r(\Delta)\}} && b_2 = |\mathcal{F}\{r(\Delta)\}|^2 \odot \Im{\mathcal{F}\{r(\Delta)\}}, \nonumber
\end{align}
where $\odot$ and $\langle \rangle_F$ are the Hadamard/element-wise product and Frobenius inner product, respectively. Since Eqn.~\ref{eqn:gradient} contains only Fourier transforms and element-wise operations, it can be calculated in $\mathcal{O}\left( N\log(N) \right)$ time.

We further accelerated L-BFGS-B by initializing $\Delta$ with a modified phase retrieval algorithm that enforces near-field constraint $r(\Delta)$ \cite{Shechtman2015}. The described algorithm was implemented in \texttt{python} using the \texttt{PyTorch} package for GPU acceleration. Using an RTX 2080, the generation of the far-field patterns for Fig.~\ref{fig:holograms} took only a few seconds. The success of the algorithm is demonstrated by the lack of a conjugate image (even in the highly undercoupled case).

\section{Effect of Substrate Reflections}
\label{appendix:substrate}

Isolated slab PhC cavities feature symmetric, bi-directional emission due to vertical reflection symmetry about the slab midplane. When placed above a reflective substrate, however, interference between the (reflected) downwards and upwards emission paths alters the cavity's radiation pattern $S(\vec{k})$ and $Q$ \cite{Kim2012}. The results are analogous to the modified spontaneous emission from a quantum emitter placed above a mirror \cite{Barnes2020}. 

\begin{figure}
\includegraphics[width=0.42\textwidth]{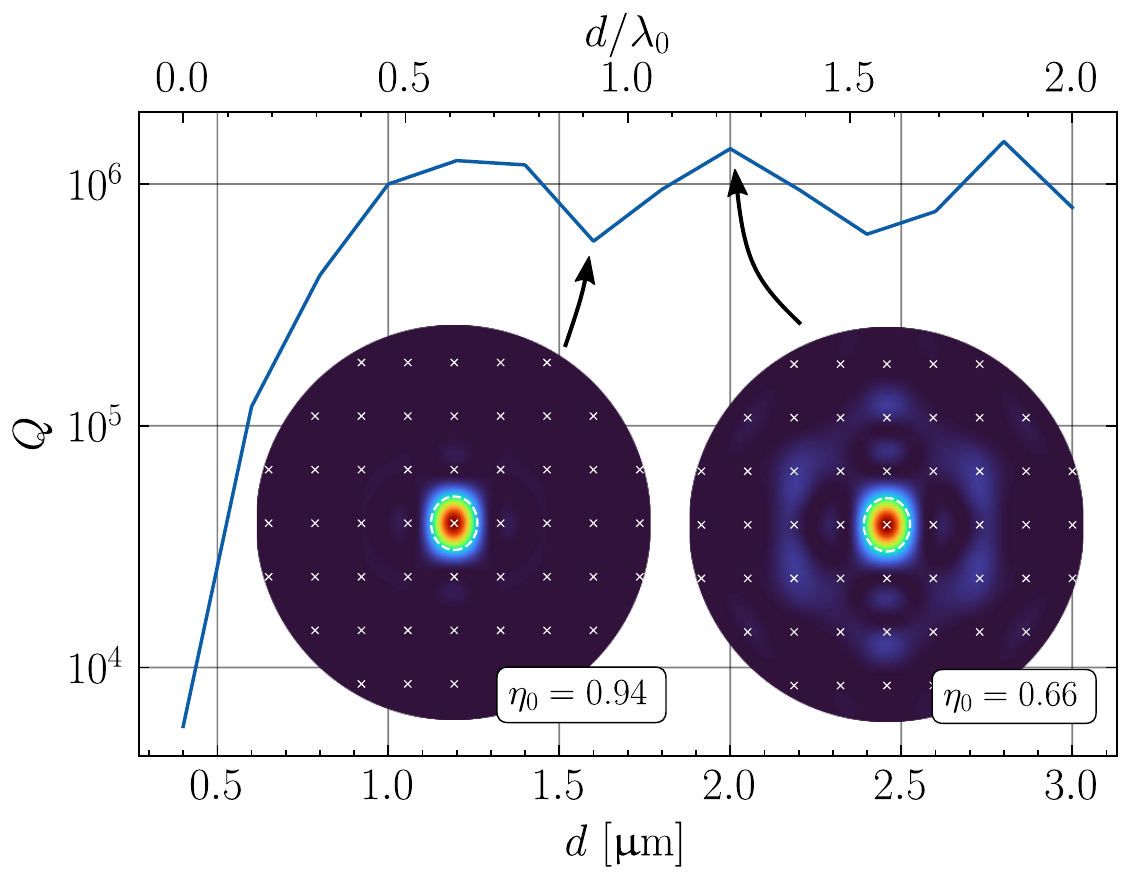}
\caption{Quality factor ($Q$) trends for an $L3$ PhC cavity (resonant wavelength $\lambda_0$) placed a distance $d$ above a silicon back-reflector. Insets show the far-field emission pattern $S(\vec{k})$ at select points.}
\centering
\label{figS:substrate}
\end{figure}

Fig.~\ref{figS:substrate} plots the FDTD-simulated quality factor trends of an inverse-designed $L3$ cavity as a function of the membrane-substrate gap spacing $d$. Since the optical thickness of the PhC slab is approximately $\lambda/2$, constructive interference at $d\approx m\lambda/2$ ($m\in 1, 2, ...$) maximizes vertical emission and minimizes $Q$. Destructive interference at $d\approx (2m-1)\lambda/4$ has the opposite effects. The resulting variation in $\eta_0$ explains the minor discrepancy between the simulated and measured $\vec{S}(k)$ in Fig.~\ref{fig:singlecav}. 

\begin{figure}
\includegraphics[width=0.3\textwidth]{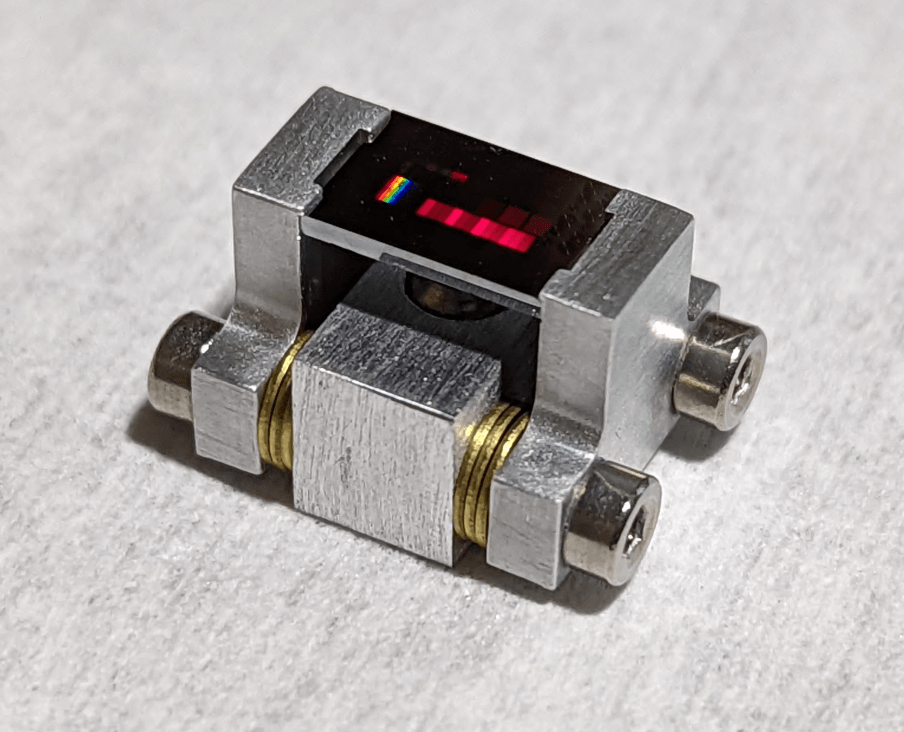}
\caption{Sample mount with set pin to flatten stress-buckled  suspended membranes.}
\centering
\label{figS:mount}
\end{figure}

\begin{figure}
\vspace{10pt}
\includegraphics[width=0.42\textwidth]{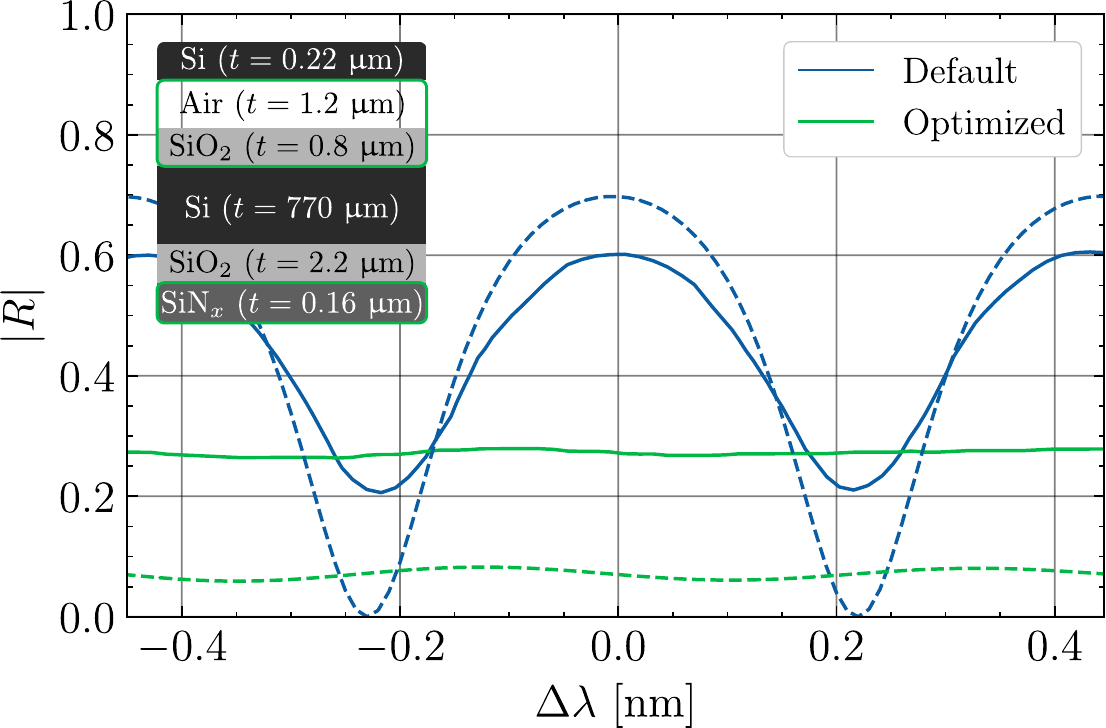}
\caption{Comparison of simulated (dashed) and measured (solid) silicon-on-insulator (SOI) sample reflection before (blue) and after (green) anti-reflection optimization for normal incidence at $\lambda=1550$ nm. Measured values were calibrated with a known reference mirror. The final layer stack of silicon (Si, $n=3.48$), oxide (SiO$_2$, $n=1.44$), and deposited silicon nitride (SiN$_x$, $n=1.90$) is shown in the inset with optimized layers highlighted in green.}
\centering
\label{figS:arc}
\end{figure}

We mitigated the impact of these effects through optimized sample preparation. Compressive stress in our silicon-on-insulator (SOI) die buckles suspended cavity arrays, yielding variations in $d$ --- and therefore cavity reflectivity --- across the membrane. As an alternative to previously proposed stress-engineered support structures \cite{Iwase2012}, we flattened the suspended arrays by mechanically bowing the die with a backside set pin in a custom sample mount (Fig.~\ref{figS:mount}). Unfortunately, the resulting (uniform) gap spacing $d=2~\upmu\text{m}$ minimizes vertical coupling at the design wavelength $\lambda_0 = 1.55~\upmu\text{m}$. We therefore added a back-side silicon nitride anti-reflection coating (ARC) on the substrate and timed the release etch to form a front-side ARC with the remaining oxide (Fig.~\ref{figS:arc}). 

\section{Far-Field Uniformity}
\label{appendix:uniformity}

\begin{figure}
\includegraphics[width=0.48\textwidth]{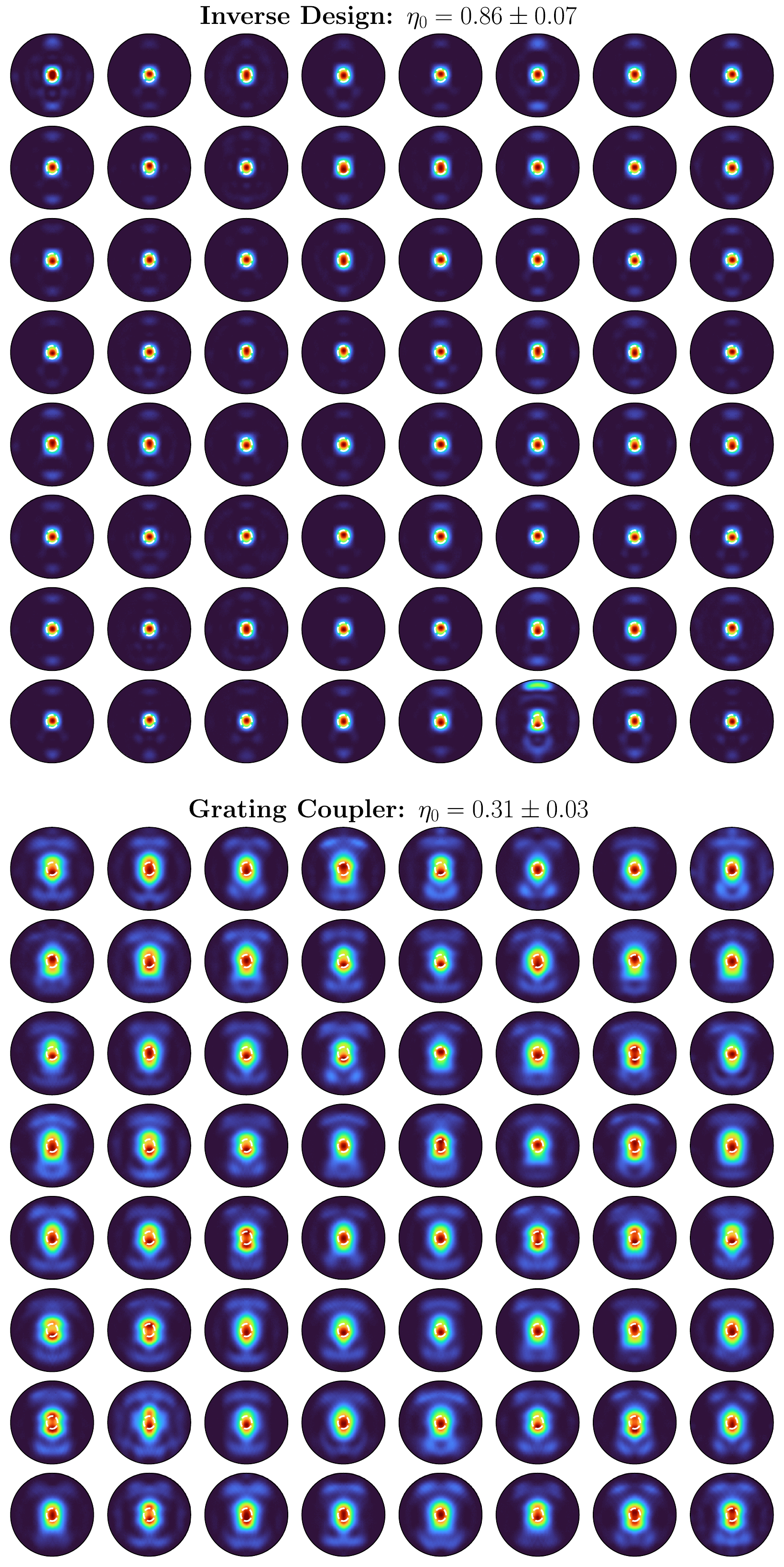}
\caption{Imaged far-field profiles $S(\vec{k})$ (over a 0.9 numerical aperture) for each device in an $8\times 8$ array of inverse designed (top) and grating-coupled (bottom) $L3$ PhC cavities. The extracted zero-order efficiencies $\eta_0$ and standard deviations are also provided.}
\centering
\label{figS:bfp-uniformity}
\end{figure}

Fig.~\ref{figS:bfp-uniformity} demonstrates the far-field uniformity characteristic of inverse-designed and grating coupled $L3$ cavity arrays. Averaged across the $8\times 8$ arrays, the former offers a $\resim 3\times$ improvement in zero-order diffraction and aperture efficiencies ($\langle \eta_0 \rangle = 0.86$, $\langle \eta_a \rangle = 0.99$).

\section{The \texttt{slm-suite} Toolbox}
\label{appendix:qpslm}

\begin{figure}
\includegraphics[width=0.4\textwidth]{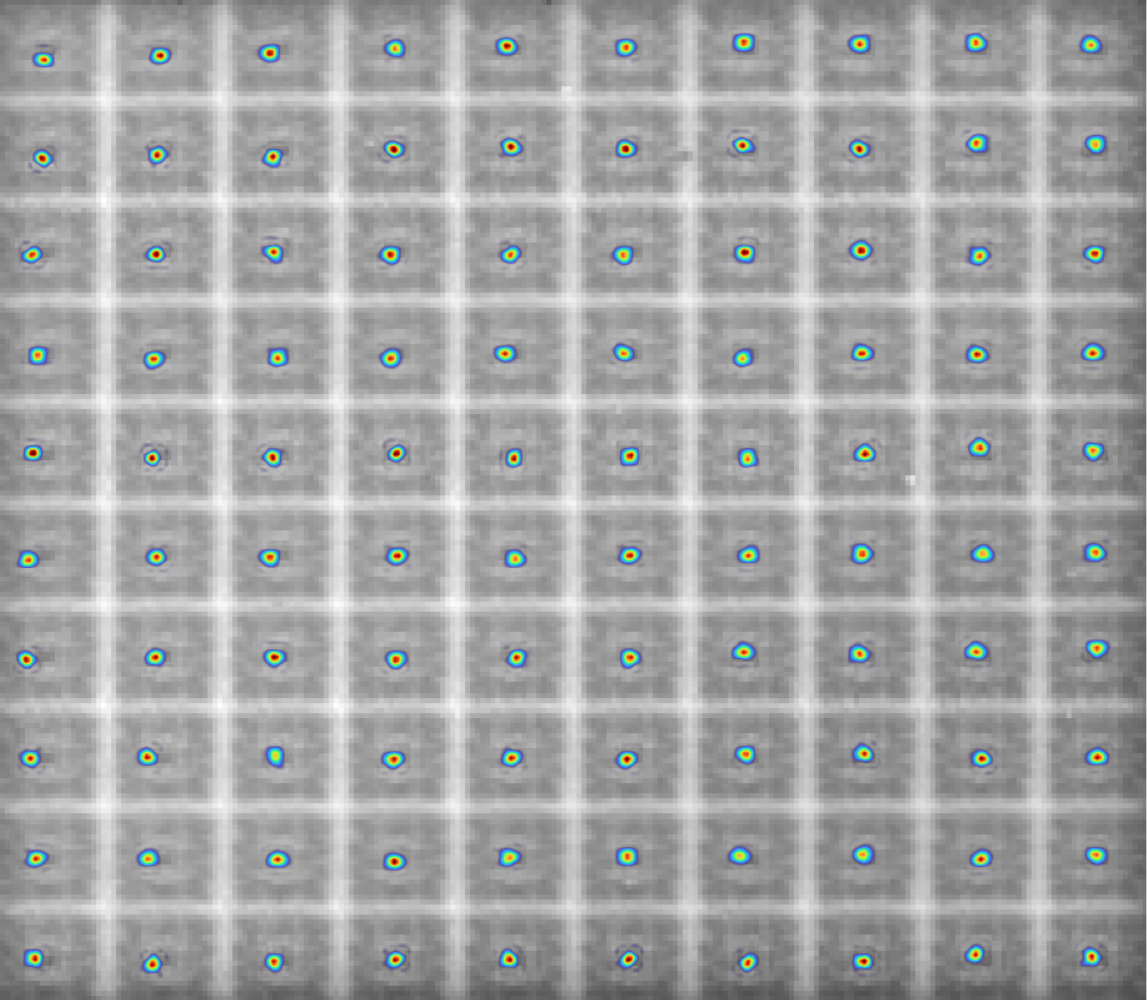}
\caption{Overlaid images of $10\times 10$ cavity (grey) and trimming spot (color) arrays demonstrating the $\ll \upmu$m placement accuracy and percent-order power uniformity of weighted Gerchberg-Saxton phase retrieval with experimental camera feedback.}
\centering
\label{figS:spotsontarget}
\end{figure}

We developed the open-source \texttt{python} package \texttt{slm-suite} to simplify the creation of high-uniformity, arbitrary-geometry optical focus arrays using various phase retrieval algorithms. The package features:
\begin{enumerate}
    \item Automated wavefront calibration routines that measure the Fourier-plane source amplitude and phase using a super-pixel interference technique to compensate for aberrations along the SLM imaging train \cite{Cizmar2010}
    \item Various graphical processing unit (GPU)-accelerated Gerchberg-Saxton (GS) algorithms that use the measured source constraints (1) to produce optimized spot array phase masks \cite{Di2007,Nogrette2014,Kim2019}
    \item Automated affine transformations between grating wave vectors applied to the SLM and image-space coordinates (i.e. camera pixels) by projecting and detecting a GS-computed spot array
    \item Camera-based feedback of measured spot amplitudes at known (calibrated) locations into phase retrieval algorithms to improve the uniformity of image-space spot arrays
    \item Automated evaluation metrics to monitor diffraction efficiency, spot amplitude and position tolerance, and spot quality.
    \item Simplified hardware interface and control.
\end{enumerate}
After calibration, high-uniformity optical foci can be generated at arbitrary image plane locations specified by the user. For example, Fig.~\ref{figS:spotsontarget} shows a $10\times 10$ spot array with $\resim 1\%$ power uniformity and sub-micron placement accuracy formed on a $10\times 10$ cavity array during the trimming procedure of Section~\ref{sec:trimming}.

\section{Parallel Laser Oxidation}
\label{appendix:trimming}

\begin{figure}
\includegraphics[width=0.4\textwidth]{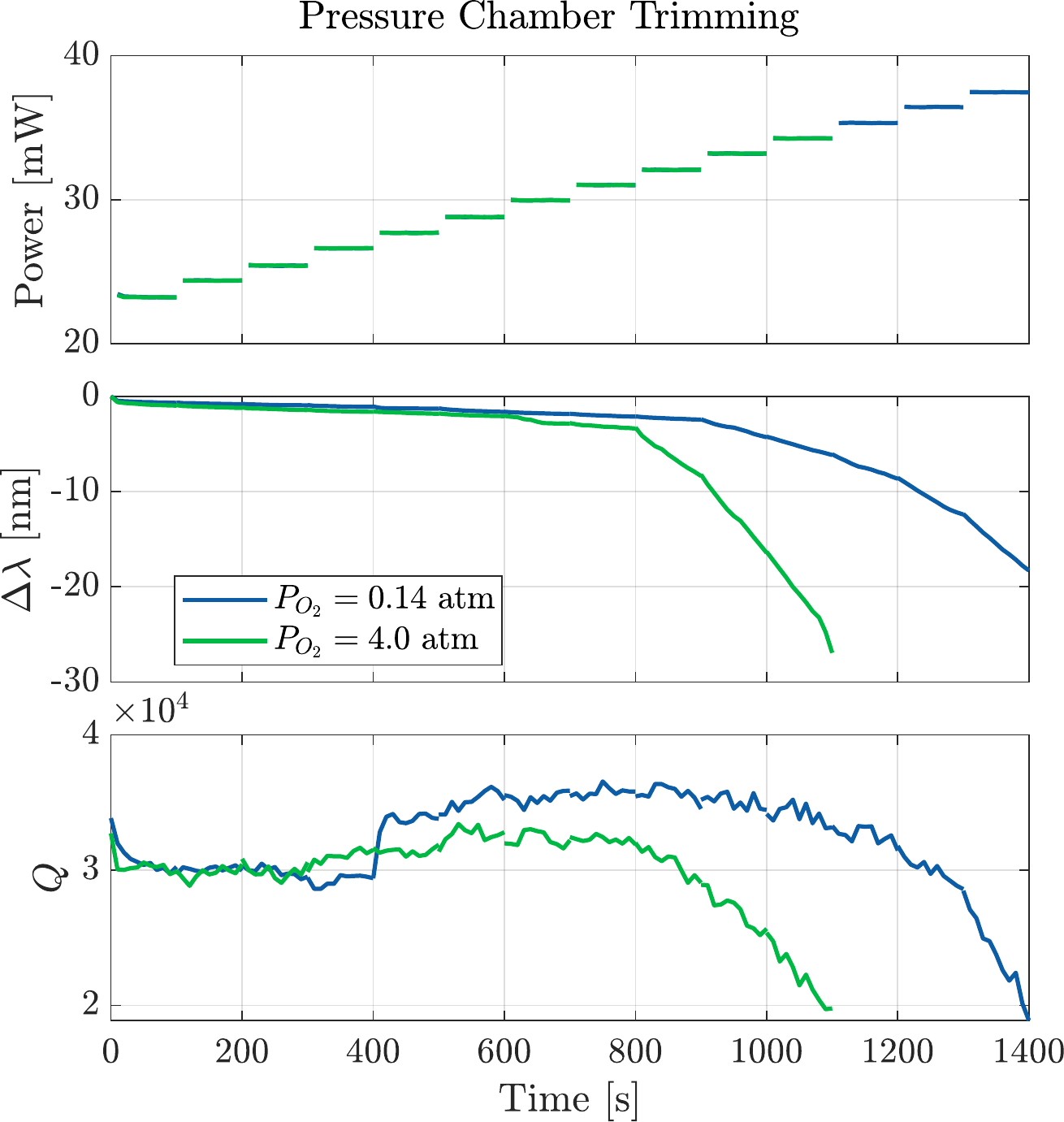}
\caption{Wavelength and quality factor trends as a function of incident visible laser power, demonstrating accelerated trimming with increased oxygen partial pressure $P_{\text{O}_2}$.}
\centering
\label{figS:hipox}
\end{figure}

As illustrated by the wavelength trends as a function of incident laser power and exposure time in Fig.~\ref{figS:hipox}, we found that laser-assisted thermal oxidation could be accelerated in a high-pressure oxygen environment. We therefore mount samples in a custom pressure chamber with a partial oxygen pressure ranging from 5 to 10 atm.

\renewcommand{\arraystretch}{1.2}
\begin{table*}
\footnotesize
\centering
\vspace{10pt}
\begin{tabular}{@{} C{5cm}C{2.5cm}C{1.5cm}C{2cm}C{2cm}C{1.5cm}C{1.5cm} @{}} \toprule
\bf Technique [Year] & \bf Cavity Type & $\bm N$ & $\bm{\Delta\lambda_0^\textbf{p-p}}$ \bf [pm] & $\bm{\langle Q \rangle }$ & \bf In situ? & \bf Parallel? \\
\midrule

\bf ``Holographic" oxidation [2022] & \bf Si PhC & \bf 64 & \bf 13 & $2\times 10^5$ & \bf Y & \bf Y \\

Germanium implantation [2021] & Si ring \cite{Jayatilleka2021} & 58 & 32 & $4\times 10^3$ & Y & N \\

Laser-annealed cladding [2020] & Si ring \cite{Biryukova2020} & 2 & 20* & $2\times 10^4$ & Y & N \\

Boron implantation [2019] & Si ring \cite{Hagan2019} & 4 & 15 & $5\times 10^3$ & Y & N \\

Electron-beam irradiation [2018] & Si PhC \cite{Han2018} & 4 & 400 & $3\times 10^5$ & N & N \\

Photo-electro-chemical etching [2017] & GaAs disk \cite{Gil2017} & 5 & 200* & $2\times 10^4$ & Y & N \\

Annealed cladding [2016] & Si ring \cite{Spector2016} & 5 & 90* & $3\times 10^3$ & Y & N \\

Ultraviolet irradiation [2014] & a-Si ring \cite{Lipka2014} & 4 & 45 & $8\times 10^3$ & Y & N \\

Post-fabrication etching [2013] & GaAs PhC \cite{Atabaki2013} & 18 & 100* & $3\times 10^4$ & N & Y \\

Photochromatic thin-film [2011] & GaAs PhC \cite{Cai2013} & 3 & 340 & $8\times 10^3$ & Y & N \\

Anodic oxidation [2006] & GaAs PhC \cite{Hennessy2006} &  2 & 100* & $5\times 10^3$ & N & N \\

\bottomrule
\end{tabular}
\caption{Comparison of microcavity array trimming techniques. Estimated values are marked with a *. $\Delta\lambda_\text{p-p}$ = peak-to-peak wavelength error; $\langle Q \rangle$ = mean quality factor.}
\label{table:trimming}
\end{table*}

\begin{figure*}
\includegraphics[width=\textwidth]{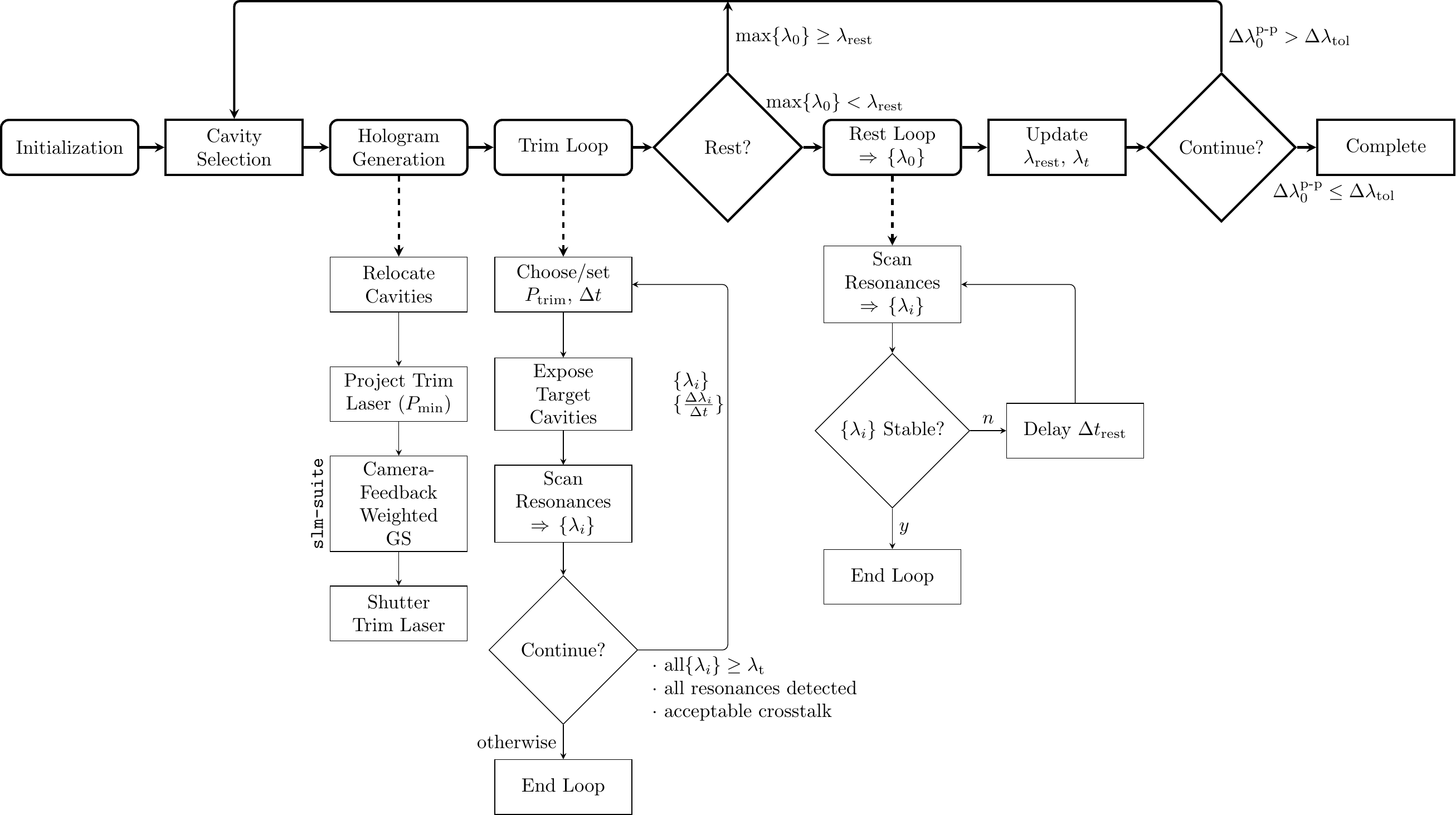}
\caption{Flowchart of the holographic trimming algorithm. Trimming holograms are formed with weighted Gerchberg-Saxton (GS) algorithms and projected onto desired cavities for duration $\Delta t$ with power $P_\text{trim}$. Alternating trimming and resonance readout periods continue until the instantaneous wavelength $\lambda_i$ of any targeted cavity blueshifts past the target wavelength $\lambda_t$. Thereafter, a new set of target cavities is selected and trimmed. This selection and trimming sub-loop continues until all resonant wavelengths $\lbrace \lambda_0 \rbrace$ are below the ``rest" wavelength $\lambda_\text{rest}$, at which point trimming is halted and the resonances are continuously monitored at readout interval $\Delta t_\text{rest}$. When the resonances are sufficiently stable (redshifting from moisture adsorption to the silicon membrane is arrested), the total ``rehydration" redshift $\Delta\lambda_0$ of each cavity is updated to better estimate the true resonant wavelength $\lambda_0 \approx \lambda_i+\Delta\lambda_0$ from the instantaneous wavelengths $\lbrace \lambda_i\rbrace$ during trimming. The entire process terminates when the peak-to-peak static resonant wavelength uniformity $\Delta\lambda_0^\text{p-p}$ drops below the desired tolerance $\Delta\lambda_\text{tol}.$}
\centering
\label{figS:trimflow}
\end{figure*}

The main loop of the trimming process (Fig.~\ref{fig:trimming}) consists of device selection, hologram setup, parallel laser oxidation, and resting intervals. The algorithm monitors two resonant wavelengths: the instantaneous wavelength $\lambda_i$ and the steady-state  wavelength $\lambda_0$. Initially $\lambda_i=\lambda_0$; however, focusing high-power ($\resim 10$ mW) visible light onto the cavity (as required to sufficiently heat the PhC membrane for thermal oxidation) causes a temporary blueshift $\Delta\lambda_0$ due to the desorption of moisture attached to hydrophilic hydroxyl surface terminations. For any target rest wavelength $\lambda_t$, we therefore trim devices to an instantaneous wavelength $\lambda_i = \lambda_t-\Delta\lambda_0$ that relaxes over $\mathcal{O}(\text{minute})$ timescales to $\lambda_i=\lambda_0=\lambda_t$  as moisture re-adsorbs to the surface. In practice, the stability and estimation of the ``overtune" $\Delta\lambda_0$ limit the uniformity and scale of the trimming process, respectively. 

After initializing the cavity locations, scanning the device resonances, and calibrating the SLM (Section~\ref{appendix:qpslm}), a spot array targeting every cavity (Fig.~\ref{figS:spotsontarget}, for example) is projected on the membrane for a short (few second) duration. Monitoring the resonances at fixed intervals $\Delta t\approx 10$~s until $\lambda_i$ stabilizes to the rest wavelength $\lambda_0$ gives an initial estimate $\Delta\lambda_0=\lambda_0-\min\lbrace \lambda_i \rbrace$ for the overtune parameter of each cavity. We also update the target wavelength $\lambda_t = \min\lbrace \lambda_0 \rbrace$ and rest wavelengths before continuing the trimming procedure. To update $\Delta\lambda_0$, we periodically conduct this same ``rest loop" when $\lambda_0$ of each cavity is below an algorithmically chosen ``checkpoint" wavelength $\lambda_\text{rest}$.

As described in Section~\ref{sec:trimming}, a subset of $N$ cavities is then selected to maximize the total possible trimming distance to $\lambda_t$. The number of targeted cavities neighboring each untargeted cavity is also limited to reduce crosstalk. A spot array is then formed to evenly distribute the trimming laser to the selected devices. After confirming that the location accuracy and power uniformity of the array are within tolerance, we alternate exposure and readout intervals to grow thermal oxide with in situ monitoring. The laser power is progressively increased to reach a desired, wavelength-uniformity-dependent trimming rate. As evidenced by Fig.~\ref{figS:hipox}, the rate is relatively power-independent until reaching a threshold power. We detect and save these threshold powers for use when selecting the initial exposure power in each trimming loop. 

The trimming sub-loop continues until the estimated $\lambda_0$ of any targeted cavity crosses $\lambda_t$. New cavities are then selected, targeted, and trimmed until a rest period is triggered. When the peak-to-peak wavelength uniformity at the end of a rest period is below the user-defined tolerance $\lambda_\text{tol}$, the process is terminated. Table~\ref{table:trimming} compares the demonstrated performance compared to other arrayed microcavity trimming techniques.

\end{document}